%% file: paper.tex
\documentclass[a4paper]{oridynamics}

\usepackage[toc,page,header]{appendix}

\usepackage{minitoc}
\usepackage{cleveref}
\usepackage{subcaption}
\usepackage{booktabs}
\usepackage[utf8]{inputenc}
\usepackage[linesnumbered,ruled,vlined]{algorithm2e}
\usepackage{amsmath}
\usepackage{float}\usepackage{fewerfloatpages}

\ifdefined\final
  \usepackage[disable]{todonotes}
\else
  \usepackage[textsize=tiny]{todonotes}
\fi
\input{macro}
\usepackage{fontawesome5}
\usepackage{tabularx}

\hypersetup{citecolor=black}
\usepackage{oridynamic_style}
\title{OriDynamics: Towards a World Model for Biomolecular Dynamics}

\affiliation[1]{LGLab Frontier AI Group}

\contribution{Full author list in Contributions}

\abstract{

\input{sections/000abstract}
}

\AtBeginDocument{\renewcommand{\bffont}{}\setstretch{1.05}\setlength{\parskip}{4pt plus 1pt minus 1pt}}
\titlespacing*{\section}{0pt}{15pt plus 3pt minus 2pt}{7pt}
\titlespacing*{\subsection}{0pt}{11pt plus 2pt minus 1pt}{5pt}
\titlespacing*{\subsubsection}{0pt}{9pt plus 2pt minus 1pt}{4pt}
\renewcommand{\checkdatalist}{%
  \begingroup
    \small\sffamily
    \urlstyle{same}
    \vspace{1mm}
    \noindent
    \begin{tabularx}{\linewidth}{
      @{}l@{\hspace{1.8em}}
      >{\footnotesize\raggedright\arraybackslash}X@{}
    }

    \href{https://lg-frontierai.github.io/homepage/oridynamics}{%
      \makebox[1.2em][c]{\faGlobe}%
      \enspace\textbf{Homepage}%
    }
    &
    \href{https://lg-frontierai.github.io/homepage/oridynamics}{%
      \textcolor{black!65}{%
        \nolinkurl{https://lg-frontierai.github.io/homepage/oridynamics}%
      }%
    }
    \\[4pt]

    \href{https://github.com/LG-FrontierAI/oridynamics}{%
      \makebox[1.2em][c]{\faGithub}%
      \enspace\textbf{GitHub}%
    }
    &
    \href{https://github.com/LG-FrontierAI/oridynamics}{%
      \textcolor{black!65}{%
        \nolinkurl{https://github.com/LG-FrontierAI/oridynamics}%
      }%
    }
    \\[4pt]

    \href{https://huggingface.co/LG-FrontierAI/oridynamics}{%
      \makebox[1.2em][c]{%
        \raisebox{-0.18em}{%
          \includegraphics[height=1.1em]{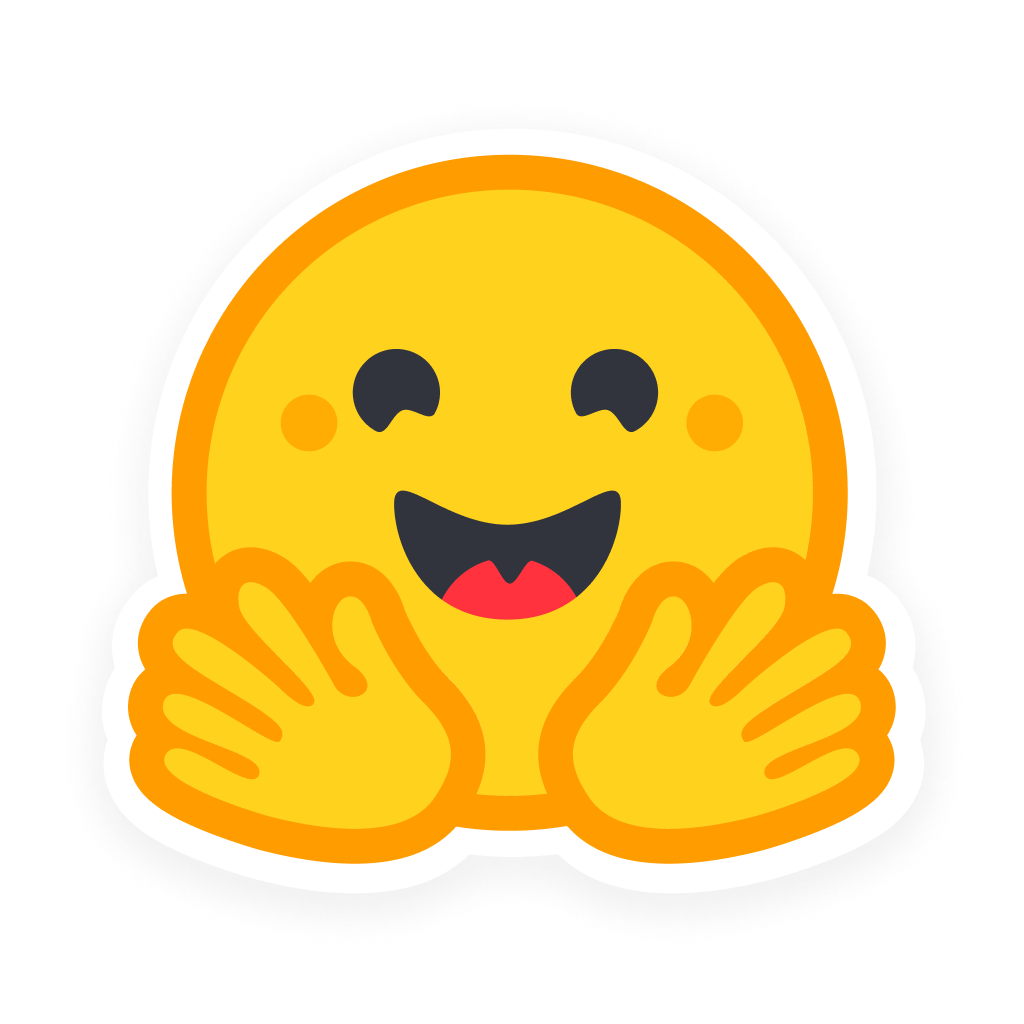}%
        }%
      }%
      \enspace\textbf{Hugging Face}%
    }
    &
    \href{https://huggingface.co/LG-FrontierAI/oridynamics}{%
      \textcolor{black!65}{%
        \nolinkurl{https://huggingface.co/LG-FrontierAI/oridynamics}%
      }%
    }

    \end{tabularx}\par
  \endgroup
}

\begin{document}

\maketitle


\input{sections/010introduction}
\input{sections/030Results}

\input{application/applications}

\FloatBarrier
\input{sections/030discussion}
\input{sections/020methods}

\input{sections/040conclusion}

\bibliographystyle{oridynamics}
{\setlength{\bibsep}{3pt plus 0.3pt}\bibliography{main}}

\clearpage

\beginsi
\setcounter{equation}{0}
\renewcommand{\theequation}{S\arabic{equation}}

\input{sections/SI}

\end{document}

%% file: macro.tex
\usepackage{natbib}
\usepackage{latexsym}

\usepackage{url}
\usepackage{amssymb}
\usepackage[utf8]{inputenc}
\usepackage{microtype}
\usepackage{booktabs}
\usepackage{pifont} 
\usepackage{multirow}
\usepackage{makecell}
\usepackage{paralist}
\usepackage{xspace}
\usepackage{color}
\usepackage{xcolor}
\usepackage{colortbl}
\usepackage{adjustbox}
\usepackage{hyperref} 
\usepackage[edges]{forest}
\usepackage{tikz} 
\usepackage{caption}
\usepackage{amsfonts}

\hypersetup{
    colorlinks,
    linkcolor={blue!80!black},
    citecolor={blue!80!black},
}
\tikzset{
    root/.style =             {align=center, text width=1cm, rounded corners=3pt, line width=0.3mm, fill=gray!10, draw=gray!80, font=\small},
    demographic/.style =         {align=center, text width=1.8cm, rounded corners=3pt, line width=0.3mm, fill=blue!10, draw=blue!80, font=\footnotesize},
    demographic_work/.style =    {align=center, text width=10cm, rounded corners=3pt, line width=0.3mm, fill=blue!10, draw=blue!0, font=\footnotesize},
    character/.style =         {align=center, text width=1.8cm, rounded corners=3pt, line width=0.3mm, fill=red!10, draw=red!80, font=\footnotesize},
    character_work/.style =    {align=center, text width=10cm, rounded corners=3pt, line width=0.3mm, fill=red!10, draw=red!0, font=\footnotesize},
    personalization/.style =           {align=center, text width=1.8cm, rounded corners=3pt, line width=0.3mm, fill=cyan!10, draw=cyan!80, font=\footnotesize},
    personalization_work/.style =      {align=center, text width=10cm, rounded corners=3pt, line width=0.3mm, fill=cyan!10, draw=cyan!0, font=\footnotesize},
    risk/.style =         {align=center, text width=1.8cm, rounded corners=3pt, line width=0.3mm, fill=orange!10, draw=orange!80, font=\footnotesize},
    risk_work/.style =    {align=center, text width=10cm, rounded corners=3pt, line width=0.3mm, fill=orange!10, draw=orange!0, font=\footnotesize},
}

\usepackage{CJK}

%% file: sections/000abstract.tex
Biomolecular recognition and regulation depend on atomic interactions that change as molecular partners fluctuate and adopt alternative conformations. Modeling these processes requires ensembles that capture both conformational diversity and coordinated atomic fluctuations between interacting molecules. Here we present OriDynamics, a unified framework for all-atom ensemble generation across protein monomers and complexes with small molecules or protein partners.
OriDynamics introduces a latent ensemble decomposition that couples broad conformational exploration in a compact structural latent space with local all-atom sampling around generated proposals. Ensemble-level supervision constrains mean positions, fluctuation amplitudes and joint coordinate covariance to capture coordinated atomic variability within and between molecular partners. Across benchmarks of proteins and their complexes, OriDynamics demonstrates superior overall performance in conformational coverage, atomic fluctuation fidelity and interaction geometry compared with leading baselines. Application studies further demonstrate recovery of alternative functional conformations and partner-dependent changes in conformational ensembles alongside sampling of configurations associated with ligand dissociation. These capabilities provide a foundation for a biomolecular world model that represents accessible molecular states and their dependence on interaction context.


%% file: sections/010introduction.tex
\section{Introduction}

Biomolecular function depends on the conformations molecules access and the fluctuations within each state \cite{yang2003,henzler2007dynamic,boehr2006,henzler2007}. Alternative conformations shape molecular recognition \cite{lange2008,hammes2009}, while changes in local flexibility and coordinated motion can influence binding and regulation \cite{frederick2007,popovych2006,vandenbedem2013}. This variability also creates opportunities for therapeutic intervention \cite{lito2016,patricelli2016}. For example, mutant-selective KRAS G12C inhibitors bind to a pocket beneath the switch-II region that was not apparent in earlier Ras structures \cite{ostrem2013}. Such findings illustrate how therapeutically relevant interactions can involve structural features absent from a single reference conformation. For drug discovery, capturing these opportunities requires representations of both accessible conformational states and the variability associated with them.

Advances in structure prediction have provided an increasingly powerful foundation for molecular discovery. AlphaFold2 and AlphaFold3 have transformed predictive accuracy and expanded modeling from protein structures to biomolecular complexes \cite{jumper2021,abramson2024}. More recently, IsoDDE and OpenDDE have reflected a broader effort to place structural modeling within workflows for molecular design and evaluation \cite{isodde,opendde}. Accurate cofolded structures, however, do not by themselves establish how a molecular system varies across alternative states or fluctuates within them \cite{cui2025beyond}. Incorporating conformational variability is particularly relevant to ensemble-based virtual screening \cite{lin2002,amaro2008,nichols2011}, cryptic-pocket discovery \cite{bowman2012,schames2004,bowman2015,oleinikovas2016} and conformation-selective inhibitor design \cite{schindler2000,xie2020,kim2023pankras}. Conformational ensembles provide a natural basis for these applications by describing accessible structural states, their associated fluctuations and their dependence on molecular context.

Molecular dynamics provides an atomistic account of conformational fluctuations and transitions, although sampling separated conformational basins remains computationally demanding \cite{shaw2010,lindorff2011}. Generative approaches have developed complementary strategies \cite{noe2019,zheng2024dig}. AlphaFlow adapts structure predictors to ensemble generation \cite{alphaflow}, while MDGen and BioKinema model temporal relationships through trajectory generation \cite{mdgen,biokinema}. BioEmu and AnewSampling target equilibrium ensemble emulation \cite{bioemu,anewsampling}, and BioEmu2 extends ensemble sampling and time-lagged propagation to broader molecular systems \cite{bioemu2}. A central challenge is to capture both the distribution of distinct conformational states and the fluctuations within them. These fluctuations carry functional information that state coverage alone cannot resolve. Changes in internal motion can alter allosteric coupling with little change in mean structure \cite{popovych2006,petit2009,tzeng2012}, while correlated displacements can connect molecular recognition sites \cite{fenwick2011}. Faithful ensemble generation therefore requires coordinated learning of conformational diversity and the magnitude, directionality and correlations of atomic fluctuations. This motivates a generative framework that couples exploration across conformational basins with conditional sampling of variability within them.

Here, we introduce OriDynamics, which combines an autoencoder, a flow module and a refine module for latent ensemble decomposition. The autoencoder learns a structural latent space through conformational encoding and reconstruction. The flow module samples latent states that decode into conformational proposals, and the refine module performs stochastic all-atom sampling conditioned on these proposals. This design couples global exploration across conformational basins through latent proposal generation with local exploitation through proposal-conditioned sampling of within-basin variability. To connect this generative organization to the target distribution, OriDynamics constrains the statistics of the final sampled ensemble. Mean-position and fluctuation-amplitude objectives supervise average atomic positions and the extent of atomic variability, while a joint coordinate covariance objective captures directional variability and correlated displacements between atoms. These objectives jointly guide the flow and refine modules, linking representative generation and conditional sampling to the collective statistics of reference ensembles.

OriDynamics provides a unified model for ensemble generation across protein monomers, protein\textendash{}ligand complexes and protein\textendash{}protein assemblies. Evaluations across four benchmark settings using public data demonstrate strong overall performance in reproducing conformational diversity, fluctuation statistics and interaction geometry. Application studies further demonstrate recovery of alternative conformations and partner-dependent apo\textendash{}holo states, with additional adaptation supporting the sampling of conformations associated with ligand dissociation. Together, these results show that OriDynamics captures diverse molecular conformations and their responses to interaction partners. This provides a foundation for a biomolecular world model that predicts accessible states and how their distributions change with molecular context.

%% file: sections/030Results.tex
\section{Results}
\label{sec:results}
\input{results/sections/overview}
\FloatBarrier
\input{results/sections/monomer}
\FloatBarrier
\input{results/sections/misato}
\FloatBarrier
\input{results/sections/dynappi}

%% file: results/sections/overview.tex
\subsection{Overview}
\label{sec:results_overview}

OriDynamics generates conformational ensembles for protein monomers and biomolecular complexes under different molecular inputs (Fig.~\ref{fig:oridynamic_overview}a). Performance across four benchmark settings is summarized by correlations in ensemble breadth and atomic flexibility (Fig.~\ref{fig:oridynamic_overview}b). The framework combines latent ensemble decomposition with supervision of mean atomic positions, fluctuation amplitudes and joint coordinate covariance (Fig.~\ref{fig:oridynamic_overview}c). Data composition and the three-stage training schedule are summarized schematically (Fig.~\ref{fig:oridynamic_overview}d,e). During inference, the flow module and autoencoder decoder generate conformational proposals, and the refine module generates the associated local ensembles through conditional all-atom diffusion (Fig.~\ref{fig:oridynamic_overview}f).

\input{results/figure_overview}

%% file: results/figure_overview.tex
\begin{figure}[!tp]
  \centering
  \begin{minipage}{\linewidth}
  \centering
  \includegraphics[width=\linewidth,keepaspectratio]{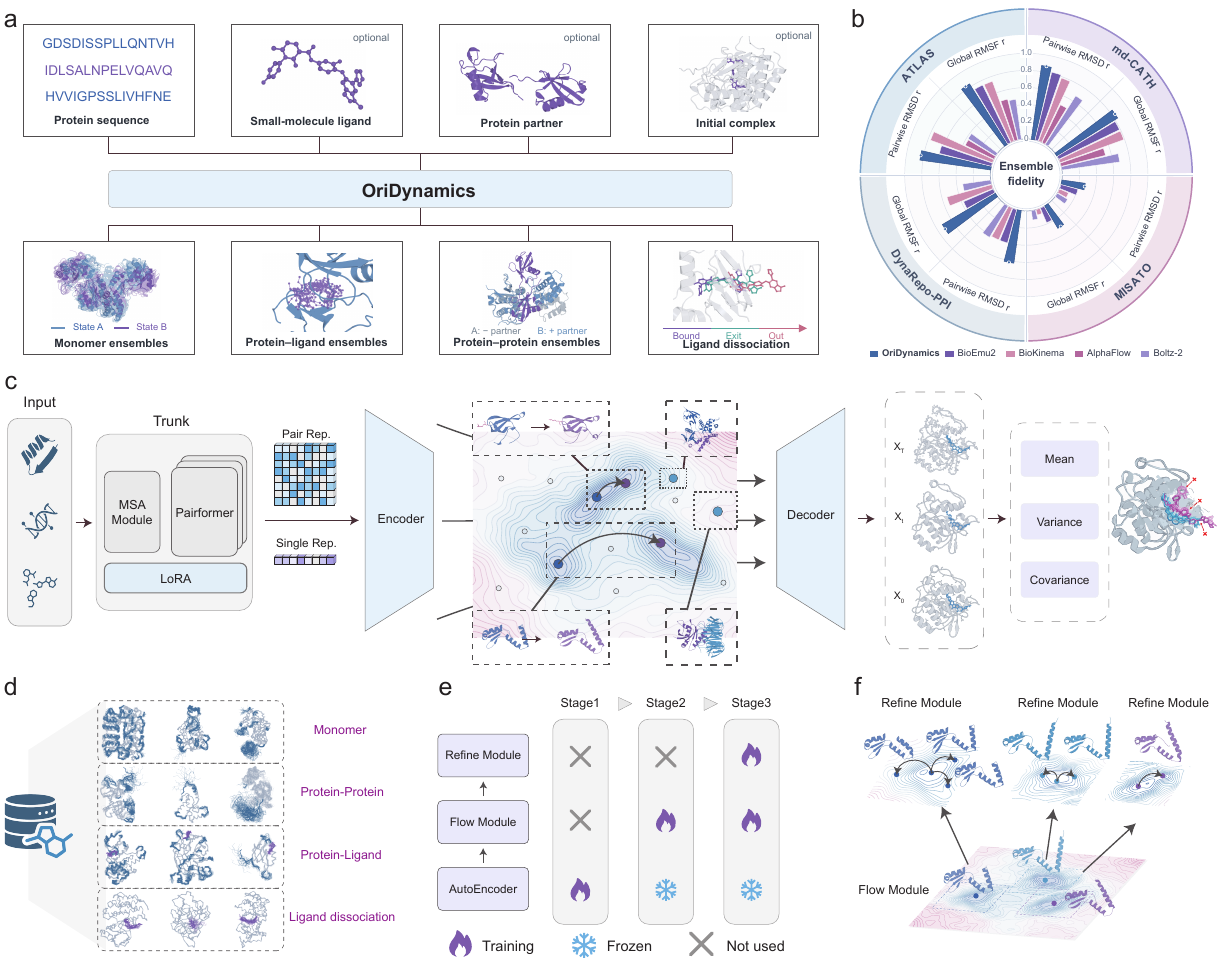}\par
  \caption[]{\input{results/captions/overview.tex}}\label{fig:oridynamic_overview}
  \end{minipage}
  \end{figure}

%% file: results/captions/overview.tex
\textbf{Overview of OriDynamics.}
\textbf{a,} Molecular inputs and supported ensemble-generation applications.
\textbf{b,} Benchmark comparison using Global RMSF $r$ and Pairwise RMSD $r$.
\textbf{c,} Latent ensemble decomposition. The flow module (FM) samples latent states that the autoencoder decodes into conformational proposals for exploration across basins, while the refine module (RM) samples the corresponding within-basin ensemble conditioned on each proposal. Ensemble-level objectives supervise the mean positions, fluctuation amplitudes and joint coordinate covariance of the final generated ensemble.
\textbf{d,} Data composition for training, adaptation and evaluation.
\textbf{e,} Three-stage training schedule. Flames, snowflakes and crosses indicate trainable, frozen and unused modules, respectively.
\textbf{f,} Schematic of the inference procedure.
\label{fig:overview}

%% file: results/sections/monomer.tex
\subsection{OriDynamics reproduces spatial flexibility and ensemble coverage}
\label{sec:results_monomer}

Protein function depends on conformational fluctuations that connect local flexibility to changes in overall shape. OriDynamics models this organization by representing an ensemble as a distribution over representative conformations, each associated with its own local all-atom fluctuations. The resulting ensembles must therefore capture both the range of accessible conformations and the organization of motion within them; broad structural coverage alone is insufficient. We evaluated these complementary aspects on the public ATLAS~\cite{atlas} and mdCATH~\cite{mdcath} test sets, comparing OriDynamics with nine other methods in ensemble breadth, residue flexibility and structural distributions (Fig.~\ref{fig:monomer_combined}).

\input{results/figure_monomer_current}

OriDynamics achieved the best performance across all three flexibility correlations on both datasets. On ATLAS, Pairwise RMSD $r$, Global RMSF $r$ and Per-target RMSF $r$ were 0.834, 0.833 and 0.892, respectively; the closest comparators were BioKinema for Pairwise RMSD $r$ and BioEmu2 for both RMSF correlations. On mdCATH, the corresponding values were 0.870, 0.836 and 0.847, compared with the strongest baseline values of 0.816 (BioEmu2), 0.790 (BioKinema) and 0.841 (BioEmu2) (Fig.~\ref{fig:monomer_combined}a,b). Thus, OriDynamics captured both differences in conformational breadth between proteins and the spatial distribution of flexibility within them. Adding the refinement module improved all three correlations on both benchmarks, including an increase in ATLAS Pairwise RMSD $r$ from 0.710 to 0.834. Providing structural context from the reference MD simulations further improved OriDynamics’s correlations across all three flexibility metrics relative to predicted conformational basins (Supplementary Figs.~\ref{fig:si_rm_ablation},\ref{fig:si_anchor_ablation}). These comparisons support the contribution of the refinement module and the initial structural context to the final ensemble.

We next examined distributional accuracy at both the level of marginal atomic distributions and that of collective conformational distributions. OriDynamics obtained median Root mean $W_2$ distance (RMWD) and Joint PCA $W_2$ distance values of 2.073 and 1.814~\AA{} on ATLAS, compared with 2.447 and 2.119~\AA{} for AlphaFlow, the closest comparator. On mdCATH, its values were 2.233 and 2.172~\AA{}, versus 2.474 and 2.219~\AA{} for BioEmu2 (Fig.~\ref{fig:monomer_combined}c,e). Together, these results are consistent with OriDynamics's central premise that representative conformations provide an effective basis for modeling the broader conformational landscape.

Performance remained broadly robust across protein lengths and secondary-structure compositions (Fig.~\ref{fig:monomer_combined}d,f). ATLAS Per-target RMSF $r$ values were approximately 0.892\textendash{}0.893 across its populated composition groups. On mdCATH, correlations were 0.863 for mixed helix/strand, 0.780 for helix-dominant and 0.752 for strand-dominant targets, although other methods led in some individual groups. Across length bins, OriDynamics's medians ranged from 0.869 to 0.917 on ATLAS and from 0.828 to 0.855 on mdCATH, without a monotonic decline with increasing length. This consistency supports the model's ability to represent flexibility across diverse monomer architectures.

The 6XRX and 5I2Y ensembles illustrate the structural basis of this agreement (Fig.~\ref{fig:monomer_combined}g,h). Their radius-of-gyration distributions closely matched MD, with $W_1$ distances of 0.52 and 0.22~\AA{}, respectively, showing recovery of global compactness across sampled conformations. Internal-distance distributions provided a more demanding view: $W_1$ errors were 6.56~\AA{} for residues 1\textendash{}508 in 6XRX and 2.84~\AA{} for residues 37\textendash{}74 in 5I2Y. OriDynamics therefore combines accurate overall extent and spatial flexibility with improved collective distributions, while particular long-range separations can remain less well reproduced. This balance is consistent with a generator that learns alternative structural representatives and then resolves local variation around them.

%% file: results/figure_monomer_current.tex
\begin{figure}[!tp]
  \centering
  \begin{minipage}{\linewidth}
  \centering
  \includegraphics[width=\linewidth,keepaspectratio]{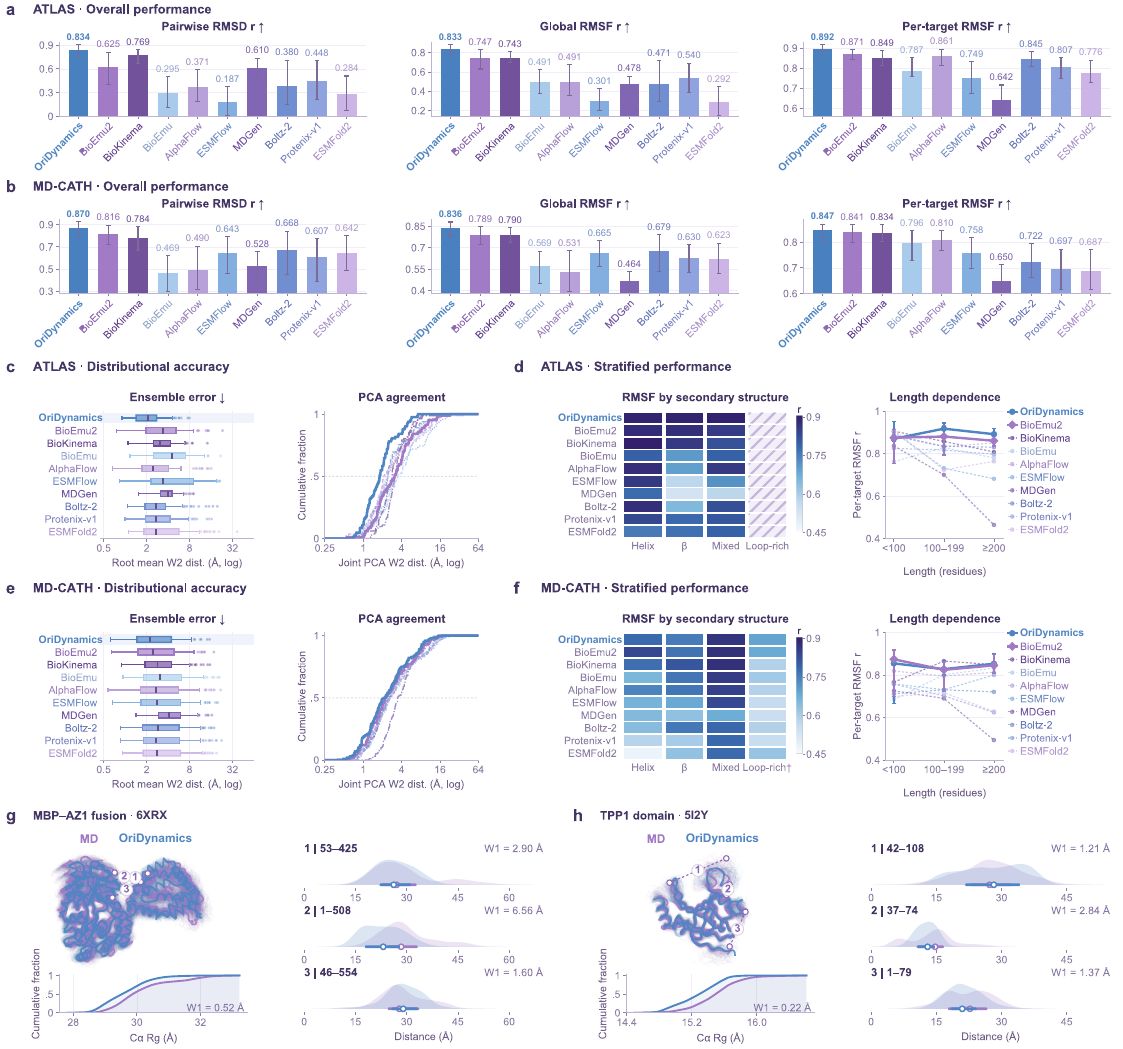}\par
  \caption[]{\input{results/captions/monomer.tex}}\label{fig:monomer_combined}
  \end{minipage}
  \end{figure}

%% file: results/captions/monomer.tex
\textbf{Monomer ensemble flexibility and structural distributions.} Methods are compared on the public ATLAS and mdCATH test sets. \textbf{a,b,} Pairwise RMSD $r$, Global RMSF $r$ and Per-target RMSF $r$. Vertical-axis ranges vary by dataset and metric and do not all start at zero; all estimates and confidence intervals are shown in full. Numbers show the point estimate for each method. \textbf{c,e,} Root mean $W_2$ distance distributions and Joint PCA $W_2$ distance cumulative distributions; lower values indicate closer agreement. Boxes show medians and interquartile ranges, with 1.5-IQR whiskers. \textbf{d,f,} Per-target RMSF $r$ by secondary structure and sequence length; heatmaps share a linear 0.45\textendash{}0.90 color scale. Hatching denotes empty groups and $\dagger$ marks sparsely populated groups. \textbf{g,h,} 6XRX and 5I2Y ensembles, residue-pair distance distributions and radii of gyration (MD, purple; OriDynamics, blue). Error bars are 95\% target-bootstrap intervals.

%% file: results/sections/misato.tex
\subsection{OriDynamics reproduces ligand and binding-pocket variability}
\label{sec:results_misato}

Ligand recognition depends on the conformational compatibility of a molecule and its binding pocket, including fluctuations that alter their contacts. An ensemble model must therefore describe the ligand and protein within a common interaction environment. To assess ligand and binding-pocket variability, we compared six methods on the public MISATO~\cite{misato} test set, progressing from ligand flexibility to pocket motion and interaction-distance distributions (Fig.~\ref{fig:misato_benchmark_overview}).

\input{results/figure_misato_current}

OriDynamics achieved the best performance across all three ligand flexibility correlations. Per-target RMSF $r$ was 0.695, compared with 0.573 for BioEmu2; Pairwise RMSD $r$ was 0.275, versus 0.205 for BioEmu2; and ligand Global RMSF $r$ was 0.303, versus 0.198 for ESMFold2, the closest comparator (Fig.~\ref{fig:misato_benchmark_overview}a\textendash{}c; 90\% confidence intervals). The gains encompass both atomic flexibility patterns within individual ligands and variation across systems. Adding the refinement module improved Per-target RMSF $r$ and Global RMSF $r$ from 0.675 to 0.695 and from 0.291 to 0.303, respectively, although Pairwise RMSD $r$ decreased from 0.337 to 0.275. MD initialization structures improved all three correlations relative to conformational basins (Supplementary Figs.~\ref{fig:si_rm_ablation},\ref{fig:si_anchor_ablation}). The benefit of the refinement module is therefore clearest for the spatial organization of ligand fluctuations.

These correlations were accompanied by fluctuation amplitudes closer to MD. Ratios of the across-target predicted and MD medians were 1.178 for ligand pairwise spread and 1.256 for RMSF amplitude; BioEmu2 and BioKinema produced larger ratios, whereas Boltz-2\cite{boltz2}, Protenix-v1\cite{protenix} and ESMFold2\cite{esmfold2} produced ratios below unity (Fig.~\ref{fig:misato_benchmark_overview}e). Agreement extended to the surrounding fixed 10~\AA{} pockets, where Per-target RMSF $r$ was 0.705, compared with 0.563 for BioEmu2. Across-target pocket Pairwise RMSD $r$ and mean-RMSF correlation were 0.437 and 0.540 (Fig.~\ref{fig:misato_benchmark_overview}e,f). OriDynamics thus reproduced flexibility on both sides of the binding interface without the pronounced overexpansion or contraction seen in several comparators.

We next examined whether these component-level improvements preserved the interactions between ligand and pocket. OriDynamics had the lowest median interaction-distance Wasserstein error in six of eight classes, including hydrogen-bond acceptors and donors, anionic interactions, $\pi$\textendash{}cation contacts, $\pi$-stacking and hydrophobic interactions (Fig.~\ref{fig:misato_benchmark_overview}d). Its errors were lower than those of BioEmu2 and BioKinema in all eight classes, whereas Boltz-2 led for cationic and cation\textendash{}$\pi$ interactions. Recovering chemically distinct distance distributions alongside atomic flexibility is consistent with sampling the ligand and protein jointly, with ensemble supervision constraining their relative coordinate variation.

The 2XZG and 5L9L cases visualize how these advantages appear within individual binding sites (Fig.~\ref{fig:misato_benchmark_overview}g,h). Pocket Per-target RMSF $r$ values of 0.598 and 0.768 locate flexible and relatively rigid regions, with mean amplitudes approximately 20\% above and 8\% below MD, respectively. Moving from these local profiles to collective pocket conformations, predicted-to-MD spread ratios in an MD-fitted PCA basis were 1.057 and 0.604. Thus, 2XZG reproduced both local variability and a projected extent close to MD, whereas 5L9L retained a narrower collective distribution despite its stronger RMSF correlation. Together, the results identify OriDynamics's strength in reproducing ligand and pocket flexibility within a shared interaction environment, while showing that accurate local amplitudes do not always imply complete recovery of collective pocket motion.

%% file: results/figure_misato_current.tex
\begin{figure}[!tp]
  \centering
  \begin{minipage}{\linewidth}
  \centering
  \includegraphics[width=\linewidth,keepaspectratio]{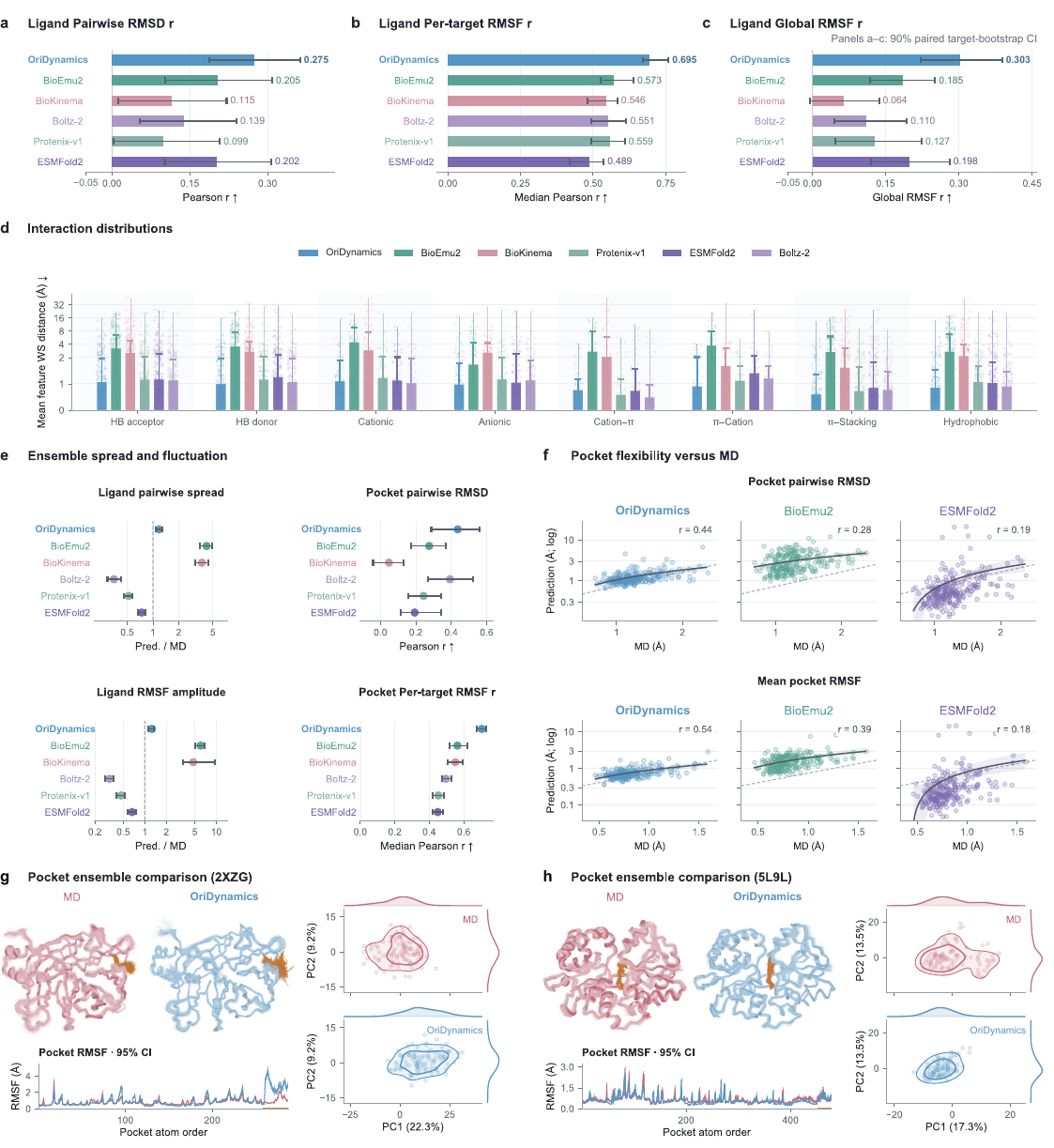}\par
  \caption[]{\input{results/captions/misato.tex}}\label{fig:misato_benchmark_overview}
  \end{minipage}
  \end{figure}

%% file: results/captions/misato.tex
\textbf{Ligand and binding-pocket variability on MISATO.} \textbf{a\textendash{}c,} Ligand Pairwise RMSD $r$, Per-target RMSF $r$ and Global RMSF $r$, with point estimates labelled and 90\% paired target-bootstrap intervals. \textbf{d,} Interaction-distance Wasserstein errors: bars show target medians, thick intervals interquartile ranges, thin whiskers observed ranges and dots individual targets. \textbf{e,} Ligand fluctuation amplitudes and pocket correlations. \textbf{f,} Predicted versus MD pocket flexibility, with regression lines and 95\% mean-fit bands. \textbf{g,h,} 2XZG and 5L9L ensembles, pocket RMSF profiles and MD-fitted PCA densities (MD, pink; prediction, blue; ligand, orange). Error bars in \textbf{e} show 95\% paired target-bootstrap intervals; profile bands describe conditional sampling uncertainty.

%% file: results/sections/dynappi.tex
\subsection{OriDynamics extends ensemble sampling to protein\textendash{}protein interfaces}
\label{sec:results_dynappi}

Protein association depends on interfaces that reorganize as their constituent chains fluctuate. Capturing this behavior requires ensembles that preserve whole-complex structure while allowing changes in local motion, solvent accessibility and interchain contacts. We therefore evaluated six methods on a 36-complex protein\textendash{}protein evaluation cohort derived from DynaRepo~\cite{dynarepo}, referred to here as DynaRepo-PPI. This evaluation extends the ligand\textendash{}pocket analysis to interactions between multiple flexible protein chains (Fig.~\ref{fig:dynarepo_benchmark_overview}).

\input{results/figure_dynappi_current}\par OriDynamics achieved the best whole-complex performance across all three flexibility correlations. Pairwise RMSD $r$ was 0.620, compared with 0.425 for Boltz-2; Per-target RMSF $r$ was 0.786, versus 0.774 for BioEmu2; and Global RMSF $r$ was 0.718, versus 0.541 for BioKinema (Fig.~\ref{fig:dynarepo_benchmark_overview}a). Adding the refinement module improved all three whole-complex correlations. MD initialization structures increased Global RMSF $r$ to 0.842, although Pairwise RMSD $r$ and Per-target RMSF $r$ decreased to 0.573 and 0.674, respectively (Supplementary Figs.~\ref{fig:si_rm_ablation},\ref{fig:si_anchor_ablation}). The shared generator therefore recovered assembly-wide flexibility, with the effect of structural context varying between global amplitudes and within-complex profiles.

At fixed 10~\AA{} interfaces, OriDynamics retained the highest Global RMSF $r$, 0.603 versus 0.505 for BioKinema, whereas BioEmu2 led in Pairwise RMSD $r$ and Per-target RMSF $r$ (Supplementary Fig.~\ref{fig:dynappi_interface_correlations}). OriDynamics also achieved the lowest mean interface RMSF absolute error, 0.801~\AA{} versus 1.221~\AA{} for BioKinema, and the lowest median Joint PCA $W_2$ distance, 2.337~\AA{} versus 2.506~\AA{} for BioEmu2. Its median Exposed residue $J$ was 0.574, compared with 0.540 for BioEmu2 (Fig.~\ref{fig:dynarepo_benchmark_overview}b,c). These results show that the model's assembly-level agreement extends to interface fluctuation amplitudes, collective distributions and changes in accessibility. Transient contacts $J$ scores, however, were 0.278 versus 0.345 for BioEmu2, indicating that recovery of individual contact rearrangements remains more challenging.

Interface cases illustrate how accurate fluctuations translate into coordinated changes in accessibility (Fig.~\ref{fig:dynarepo_benchmark_overview}b). In A00C7 and A00CE, Exposed MI matrix $\rho$ values were 0.625 and 0.628. A00C7 shared 19 dynamically exposed residues with MD, with a Jaccard score of 0.792, while A00CE recovered the same exposed-residue set. The corresponding information maps capture relationships among exposure states beyond identifying the participating residues. Interface Per-target RMSF $r$ values of 0.793 and 0.899 for A00CC and A00C5 further demonstrate recovery of the spatial pattern of motion along the interaction surface.

Contact maps provide the next level of detail by resolving how these motions alter interchain connections (Fig.~\ref{fig:dynarepo_benchmark_overview}c). A00BU and A00CI achieved Transient contacts $J$ scores of 0.621 and 0.400, and Weak contacts $J$ scores of 0.591 and 0.568. Shared MD-PCA projections for A00BS and A00CU showed overlapping predicted and reference regions, alongside differences in density and spread. The cases illustrate how joint sampling can recover interface fluctuations, exposure changes and alternative contact arrangements within the same ensemble. Overall, OriDynamics is strongest in the continuous description of multichain flexibility and interface accessibility; its lower aggregate transient-contact overlap identifies a remaining gap in resolving specific interchain rearrangements. This distinction clarifies the capabilities of a framework that supervises both fluctuation amplitudes and joint coordinate variation.

%% file: results/figure_dynappi_current.tex
\begin{figure}[!tp]
  \centering
  \begin{minipage}{\linewidth}
  \centering
  \includegraphics[width=\linewidth,keepaspectratio]{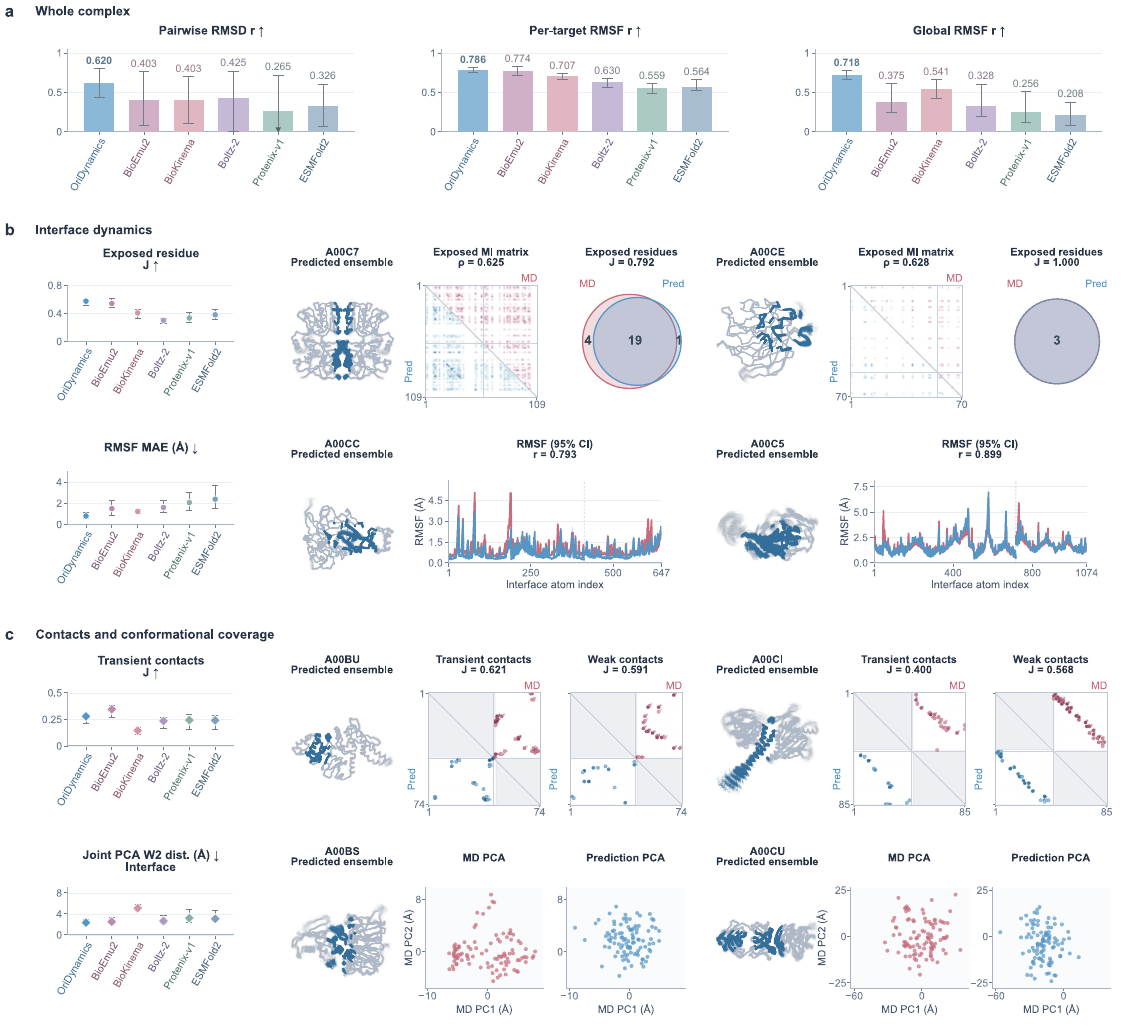}\par
  \caption[]{\input{results/captions/dynappi.tex}}\label{fig:dynarepo_benchmark_overview}
  \end{minipage}
  \end{figure}

%% file: results/captions/dynappi.tex
\textbf{Protein\textendash{}protein ensemble variability on DynaRepo-PPI.} \textbf{a,} Whole-complex Pairwise RMSD $r$, Per-target RMSF $r$ and Global RMSF $r$, with point estimates labelled. \textbf{b,} Interface exposure and RMSF errors, exposure-information maps and fluctuation profiles. \textbf{c,} Contact recovery and conformational coverage, illustrated by contact maps and shared MD-PCA projections. MD is pink and predictions blue; matrix upper/lower triangles show MD/prediction. Summary bars are 95\% target-bootstrap intervals and profile bands are pointwise frame-bootstrap intervals. RMSF correlations use heavy atoms; amplitude-error, contact and PCA metrics use C$\alpha$ atoms.

%% file: application/applications.tex
\FloatBarrier
\section{Applications}
\label{sec:applications}

The ensemble benchmarks establish correspondence to MD reference distributions. The following applications examine how this translates to function-related structural variation: recovery of alternative experimental states, responses to changes in molecular partners, and coverage of ligand-dissociation-associated configurations. These studies connect representative-state sampling and conditional fluctuations to distinct biological settings.

\input{application/sections/functional_states}
\input{application/sections/dd13m}

%% file: application/sections/functional_states.tex
\subsection{OriDynamics samples alternative states and partner-dependent responses}
\label{sec:app_functional_states}
\label{sec:app_conformation_change}
\label{sec:app_induced_fit}

Access to alternative conformations enables proteins to expose binding sites, rearrange domains and accommodate molecular partners. A useful ensemble generator must recover these states and respond when the interaction environment changes. We assessed these complementary capabilities using the public ConforMix~\cite{oridata_conformix2025} and ProMiSE~\cite{oridata_promise2026} benchmarks: the former tests recovery of alternative experimental endpoints from a common input, whereas the latter tests structural responses to the addition of a ligand or protein partner. These evaluations connect OriDynamics's generation of alternative representatives and context-conditioned all-atom sampling to functional conformational change.

On ConforMix, OriDynamics achieved the best overall TM-score dual-state success among seven methods, recovering both endpoints in 64.8\% of cases, compared with 60.4\% for BioEmu and 48.4\% for BioEmu2 (Fig.~\ref{fig:app_functional_states}a). OriDynamics also achieved the highest overall success under the RMSD criterion, with success rates of 40.7\%, 37.4\% and 20.9\%, respectively (Supplementary Fig.~\ref{fig:app_functional_states_supp}a). The consistent advantage under both criteria supports OriDynamics's ability to recover distinct conformational states. This outcome aligns with the central idea of latent ensemble decomposition: representative conformations provide distinct structural contexts for ensemble generation, organizing sampling across multiple regions of conformational space that together encompass alternative experimental endpoints.

\input{application/figure_functional_states_current}

To visualize the sampled distributions more clearly, we used separate 1,000-structure ensembles for the case studies. Maltose-binding protein illustrates recovery of a binding-cleft rearrangement: minimum RMSDs to States A and B were 0.63 and 0.70~\AA{}, respectively (Fig.~\ref{fig:app_functional_states}c). The W62\textendash{}W340 C$\alpha$ separation changed from 19.2 to 14.9~\AA{} between experimental endpoints and from 19.0 to 15.5~\AA{} between sampled representatives. Bs pyrophosphatase extended this behavior to domain motion, reaching 0.93 and 0.86~\AA{} to the two endpoints (Fig.~\ref{fig:app_functional_states}d). In both cases, the sampled ensembles reached the alternative experimental conformations, while the residue profiles characterized variation within the corresponding structural groups. The recovered binding-cleft and domain arrangements were thus represented by families of related structures. This organization reflects OriDynamics's use of representative conformations to give structural context to ensemble variability, linking experimentally recognizable arrangements to the diversity of structures associated with each.

More localized changes provide a stricter test of the sampled conformations. SpPaaI reached endpoint RMSDs of 0.87 and 0.81~\AA{}; the local overlay compares the sampled and reference configurations around residues 7\textendash{}14 (Fig.~\ref{fig:app_functional_states}e). BbFpn reached endpoint RMSDs of 1.13 and 0.97~\AA{}; the aligned views of residues 257\textendash{}276 reveal the segment displacement underlying the global scores (Fig.~\ref{fig:app_functional_states}f). Together, these views complement the global endpoint scores by resolving where sampled conformations agree with the references and where local differences remain.

We next tested whether alternative-state sampling was sensitive to molecular partners. Following the cluster-averaged reporting used by ProMiSE, which prevents heavily represented protein groups from dominating the comparison, OriDynamics achieved the highest double-state success at 54.7\%, followed by ESMFold2 at 46.9\%, BioEmu2 at 45.4\%, Protenix-v1 at 43.5\% and Boltz-2 at 39.0\% (Fig.~\ref{fig:app_functional_states}b). OriDynamics's outcomes comprised 54.7\% both-state recovery, 14.1\% apo-only, 23.3\% holo-only and 7.9\% neither. Its lower holo-only fraction than ESMFold2, Protenix-v1 and Boltz-2 indicates that stronger dual-state recovery arose from recovering apo conformations alongside bound states. The cluster-weighted advantage persisted across low-, intermediate- and high-displacement groups (defined in Appendix~\ref{sec:eval_promise_groups}), with success rates of 80.5\%, 58.3\% and 25.9\%. Equal pair weighting instead favored ESMFold2 (49.6\% versus 45.4\%), so the leading result is specific to the cluster-weighted comparison. Removing the largest two clusters reversed the pair-weighted difference in a sensitivity analysis (Supplementary Fig.~\ref{fig:app_functional_states_supp}b\textendash{}e).

CMP kinase visualizes this responsiveness to a ligand (Fig.~\ref{fig:app_functional_states}g). Without a partner, the sampled distribution favored the apo-associated region; adding ara-CMP shifted it toward the holo endpoint. The closest representatives reached apo and holo C$\alpha$ RMSDs of 1.01 and 0.93~\AA{}, respectively, with the highlighted binding region locating the structural response. RimM demonstrates the corresponding behavior for a protein partner: ensembles generated without and with ribosomal protein S19 reached endpoint RMSDs of 1.39 and 0.97~\AA{} (Fig.~\ref{fig:app_functional_states}h). The shifts in RMSD distributions and residue-deviation profiles show that partner conditioning changes the sampled ensemble as well as its closest endpoint match. Analyses of response direction and localization, together with the xylose-binding protein and LotA OTU1 examples, extend this picture to additional interaction contexts; sampling-budget comparisons assess its dependence on ensemble size (Supplementary Fig.~\ref{fig:app_functional_states_supp}d,f\textendash{}h).

Together, these results show that OriDynamics covers alternative conformations from a fixed molecular input and adapts its ensembles to molecular partners, from cleft and domain rearrangements to ligand- and protein-dependent responses. Mapping FM structural representatives onto RM-derived distributions provides a complementary view of this organization (Supplementary Fig.~\ref{fig:app_fm_rm_landscapes}). This illustrates the rationale of latent ensemble decomposition: alternative representatives provide structural contexts for conditional all-atom generation, connecting diversity among conformational arrangements with the configurations associated with each.

%% file: application/figure_functional_states_current.tex
\begin{figure}[!tp]
\centering
\begin{minipage}{\linewidth}
\centering
\includegraphics[width=\linewidth,keepaspectratio]{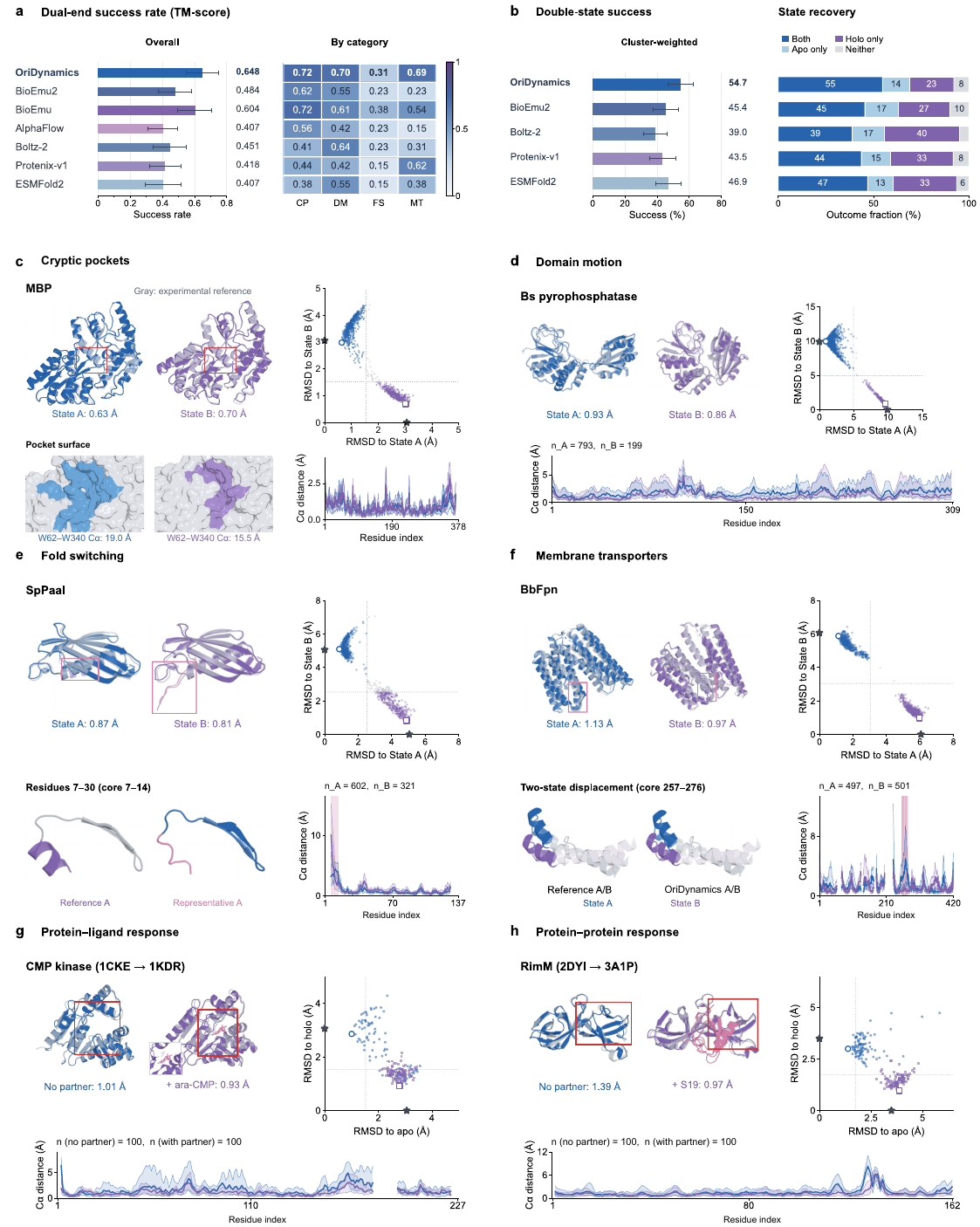}\par
\caption[]{\input{application/captions/functional_states}}
\label{fig:app_functional_states}
\label{fig:app_conformation_change}
\label{fig:app_induced_fit}
\end{minipage}
\end{figure}

%% file: application/captions/functional_states.tex
\textbf{Alternative-state coverage and partner-dependent responses.} \textbf{a,} ConforMix TM-score dual-end success. \textbf{b,} ProMiSE cluster-weighted success and recovery composition. \textbf{c\textendash{}f,} ConforMix case visualizations; blue/purple identify A/B, with experimental references overlaid in gray. Close-ups: MBP pocket surfaces (c), A/B pairs (f), reference/representative A (e, purple/pink). MBP labels give W62\textendash{}W340 C$\alpha$ distances. \textbf{g,h,} CMP kinase and RimM without/with partners (blue/purple); red boxes highlight binding-related regions. Stars/outlined symbols denote references/minimum-RMSD representatives. Dotted guides: half the reference\textendash{}reference RMSD. $n_A$ and $n_B$ denote conformer counts in the corresponding state-associated groups. Profiles show median C$\alpha$ deviations and 10th\textendash{}90th percentiles. Error bars: 95\% target (a) and dependency-block (b) bootstrap intervals.

%% file: application/sections/dd13m.tex
\subsection{OriDynamics expands ligand-pose coverage and recovers interaction distributions}
\label{sec:app_dd13m}

Ligand dissociation involves substantial changes in position, orientation and protein contacts, making it a demanding test of conformational sampling beyond bound-state fluctuations. We therefore evaluated the OriDynamics model adapted on DD-13M~\cite{oridata_dd13m2026} against BioKinema and UnbindingFlow on the public test partition, using 300 predictions per method per complex and the corresponding weighted MD ensembles (Fig.~\ref{fig:app_dd13m}). We assessed pose coverage and precision alongside ligand\textendash{}protein interaction distributions to determine how faithfully the generated ensembles represented the structural diversity associated with dissociation.

\input{application/figure_dd13m_current}

OriDynamics achieved the highest reference-pose coverage at a 2~\AA{} threshold, reaching 66.0\% coverage and 66.3\% precision, compared with 52.3\% and 95.5\% for BioKinema and 40.6\% and 9.2\% for UnbindingFlow (Fig.~\ref{fig:app_dd13m}a). At 5~\AA{}, OriDynamics reached 94.3\% coverage and 97.9\% precision. The model therefore covered a broader range of MD poses, whereas BioKinema concentrated predictions more closely around reference configurations at the stricter threshold. State-resolved analysis localized the coverage advantage to displaced configurations: OriDynamics's 2~\AA{} coverage across S0\textendash{}S3 (defined in Appendix~\ref{sec:eval_dd13_states}) was 96.4\%, 81.2\%, 52.4\% and 13.3\%, respectively. In S2 and S3, BioKinema reached 7.6\% and 0.1\%, and UnbindingFlow 18.3\% and 6.3\% (Fig.~\ref{fig:app_dd13m}b). The larger gains in displaced states indicate that OriDynamics explores configurations beyond the bound region, consistent with generating alternative representatives through latent flow sampling. Coverage remained lower for S3, identifying the most displaced configurations as the hardest regime within this comparison.

Broader coverage was accompanied by closer interaction and displacement distributions (Fig.~\ref{fig:app_dd13m}c). Mean contact-probability errors were 0.0207 for OriDynamics, 0.0271 for BioKinema and 0.0400 for UnbindingFlow. Native-contact-fraction $W_1$ errors were 0.201, 0.307 and 0.411, and centroid-displacement errors were 1.43, 2.35 and 5.58~\AA{}, respectively. OriDynamics also achieved the highest mean ligand Per-target RMSF $r$, 0.467 versus 0.341 and 0.248 (Fig.~\ref{fig:app_dd13m}d). These improvements show that broader pose coverage is accompanied by closer contact and fluctuation distributions. This combination aligns with the purpose of the refinement module and joint ensemble supervision: local all-atom sampling is conditioned on complete molecular representatives, and the final ensembles are constrained in both fluctuation amplitude and coordinated atomic displacement.

The cases resolve these advantages at the ligand and pocket levels. For 4EEV and 2GQG, ligand Per-target RMSF $r$ values were 0.851 and 0.844, with predicted-to-MD amplitude ratios of 0.96 and 0.83 (Fig.~\ref{fig:app_dd13m}e), illustrating recovery of heterogeneous atomic flexibility without excessive expansion. The 1T31 pocket exposure-information maps correlated at 0.48. For 4GU6 ligand 44, contact-probability MAE was 0.092 over the eight displayed variable contacts and 0.113 across all variable contacts (Fig.~\ref{fig:app_dd13m}f). The ensembles therefore captured aspects of ligand motion together with changes in pocket exposure and contact retention.

Collective ligand distributions provided a final test of this multiscale agreement. Mean ligand PCA $W_2$ distances were 2.78~\AA{} for OriDynamics, 3.57~\AA{} for BioKinema and 6.34~\AA{} for UnbindingFlow; OriDynamics had lower errors in 82.1\% and 96.4\% of paired comparisons, respectively (Fig.~\ref{fig:app_dd13m}g). For 4GU6 ligands 39 and 45, the largest nearest-prediction RMSD across the MD state medoids was 1.18 and 1.30~\AA{}, while ensemble $W_2$ errors were 3.58 and 2.32~\AA{} (Fig.~\ref{fig:app_dd13m}h,i). These examples illustrate the two complementary outcomes targeted by latent ensemble decomposition: recovering distinct representative poses and describing the configurations around them. Together with the coverage, contact and RMSF comparisons, they show that the adapted OriDynamics model combines broad ligand-pose exploration with a detailed description of local molecular variation. The advantage therefore spans both the range of sampled configurations and their interaction statistics, supporting an ensemble-level approach that links representative ligand\textendash{}protein arrangements with the structural variability associated with each.

%% file: application/figure_dd13m_current.tex
\begin{figure}[!tp]
  \centering
  \begin{minipage}{\linewidth}
  \centering
  \includegraphics[width=\linewidth,keepaspectratio]{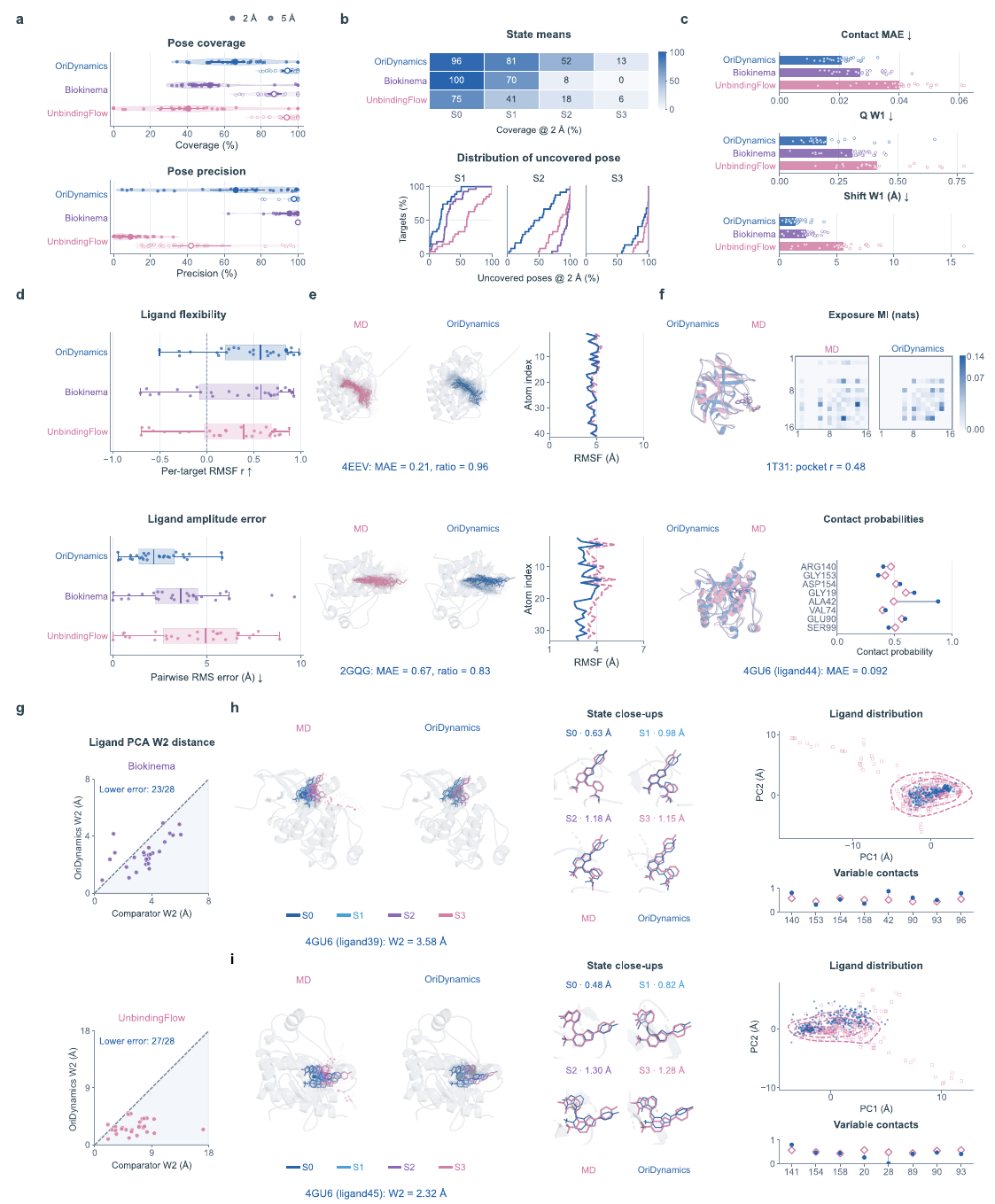}\par
  \caption[]{\input{application/captions/dd13m.tex}}\label{fig:app_dd13m}
  \end{minipage}
  \end{figure}

%% file: application/captions/dd13m.tex
\textbf{Ligand ensembles on DD-13M.} \textbf{a,b,} Pose coverage/precision and MD-defined state coverage. \textbf{c,d,} Contact, displacement and ligand-fluctuation errors. Dots denote complexes; bars are descriptive 95\% sequence-component bootstrap intervals. \textbf{e,f,} RMSF, pocket exposure-information and contact comparisons. \textbf{g,} Paired ligand PCA $W_2$ distances. \textbf{h,i,} 4GU6 ligands 39/45: MD state representatives, nearest predictions and ensemble distributions. MD is pink and OriDynamics blue in paired views; state colors link representatives and predictions.

%% file: sections/030discussion.tex
\section{Discussion}

OriDynamics learns all-atom biomolecular ensembles through latent ensemble decomposition. An autoencoder establishes a structural latent representation, a flow module generates conformational proposals, and a refine module produces all-atom ensembles conditioned on these proposals. This organization connects exploration across conformational basins with the description of variability within them. Ensemble-level supervision of mean atomic positions, fluctuation amplitudes and joint coordinate covariance complements the generative design by constraining the collective properties of the final conformations. Together, these components make both the discovery of alternative structural arrangements and the organization of atomic fluctuations explicit targets of learning.

The results provide complementary evidence for this design. Recovery of alternative experimental states, including domain rearrangements and binding-cleft motions, supports the capacity for substantial conformational exploration. Agreement with reference fluctuation and interaction distributions further supports the statistical description of the generated ensembles. Their combination suggests that structural diversity and coordinated atomic variability can be addressed within a unified model across monomers and molecular complexes. The partner-dependent responses further connect internal conformational adjustment with changes in the surrounding molecular entities. This supports a broader description in which a molecular system responds both to changes in its own structural state and to changes in its interaction environment.

For drug discovery, a promising application is the investigation of rare or difficult-to-sample conformations that reveal new opportunities for molecular intervention. Domain rearrangements, pocket opening and changes in interface organization may expose sites that are absent from commonly observed structures. Characterizing these states could expand the range of target conformations available for ligand discovery, while analysis of their associated fluctuations could guide hypotheses about coupling between newly exposed sites and distant functional regions. Ligand-binding and perturbation experiments could then establish whether these structural relationships support regulation. This creates a prospective route from conformational exploration to cryptic-pocket identification and the investigation of candidate allosteric sites, with potential applications in conformation-selective recognition and molecular design.


By representing conformational variability and its dependence on interaction partners, OriDynamics provides a foundation for a biomolecular world model. Extending the framework to predict evolution over physical time would require temporal modeling and validation against kinetic measurements. Broader coverage of flexible antibody and antigen interfaces, multidomain assemblies and complexes containing disordered regions will also require more diverse conformational data and experimental constraints. These resources could improve local geometry and intermolecular interactions while extending the range of accessible states. Modeling the effects of solution chemistry, temperature and membrane environments would further enable predictions under different physical conditions. Integration with large language models could help connect ensemble predictions with biological knowledge to formulate mechanistic hypotheses and select informative experiments. Physical modeling and experimental feedback could then test these hypotheses and guide further sampling. Together, these developments would advance biomolecular world models that connect molecular states, environmental responses and experimentally testable predictions.

%% file: sections/020methods.tex
\section{Methods}
\label{sec:methods}
\input{sections/020data}
\input{sections/020architecture}
\input{sections/020evaluation}

%% file: sections/020data.tex
\subsection{Data}
\label{sec:methods_data}

We assembled biomolecular conformational data covering protein monomers and domains, protein\textendash{}ligand complexes, and protein\textendash{}protein assemblies for model development and evaluation. Data preparation removed duplicate targets and target-level redundancy, maintained molecular and atom identities through source-specific mappings, and established consistent atom correspondence between reference structures and ensemble conformations. Invalid coordinates were excluded from supervision using validity masks, while coordinate preprocessing preserved the relative placement of interacting partners. Following preprocessing, the data were organized under a common partitioning framework, and the training ensembles were combined for joint training of a single model.

Evaluation used public test partitions and benchmark cohorts spanning monomer flexibility, ligand and pocket variability, protein\textendash{}protein assemblies (DynaRepo-PPI, our evaluation cohort derived from DynaRepo), alternative experimental states (ConforMix), and partner-dependent responses (ProMiSE). Ligand-dissociation experiments used a separate adaptation and evaluation on DD-13M. Each dataset underwent target-level deduplication and redundancy control, with complete molecular targets kept separate across training, validation and testing. The application cohorts shared no complete molecular targets with the MD training corpus. Appendix~\ref{sec:supp_data_review} summarizes the data sources, molecular correspondence and target-level overlap controls.

%% file: sections/020architecture.tex
\subsection{Model Architecture}
\label{sec:methods_architecture}

OriDynamics combines an autoencoder (AE), a latent flow module (FM) and a refine module (RM) to implement latent ensemble decomposition. The FM and AE decoder jointly generate conformational proposals for exploration across conformational basins, while the RM generates local ensembles conditioned on these proposals. Molecular identity, chemical features and available evolutionary information are encoded by an AlphaFold 3\textendash{}like trunk~\cite{abramson2024}, whose representations condition these modules.

The AE encodes all-atom conformations into token-level latent representations and learns to reconstruct their atomic geometry. The FM generates new states in this learned latent space, and the AE decoder maps these states to all-atom conformational proposals. With molecular and structural inputs held fixed, proposal generation is written as

\begin{equation}
h=\mathcal F_{\phi}(\epsilon_z),\quad \epsilon_z\sim\mathcal N(0,I),\quad X_{\mathrm{prop}}=\mathcal D_{\psi}(h).
\label{eq:main_generative_pipeline}
\end{equation}
Here, $\mathcal F_{\phi}$ denotes latent flow sampling with inverse normalization, and $\mathcal D_{\psi}$ is the AE decoder. Conditioning on the fixed inputs is omitted from both sampling equations for clarity.

The RM is adapted from a diffusion module. Conditioned on the decoded proposal $X_{\mathrm{prop}}$ in Eq.~\ref{eq:main_generative_pipeline}, it directly generates all-atom conformations:
\begin{equation}
X=\mathcal R_{\omega}\!\left(X_{\mathrm{prop}}\right).
\label{eq:main_refinement_sampling}
\end{equation}
Here, $\mathcal R_{\omega}$ denotes proposal-conditioned diffusion generation. The RM improves molecular geometry and generates the local conformational ensemble within the basin represented by the proposal. Together, latent proposal generation and all-atom diffusion generation support exploration across basins and description of within-basin variability. Representation and network details are provided in Appendix~\ref{sec:si_architecture}, and sampling procedures are described in Appendix~\ref{sec:si_inference}.

\subsection{Training and Loss Functions}
\label{sec:methods_training}

Training follows three stages. First, the AE is pretrained on molecular dynamics (MD) conformations using coordinate and pair-distance reconstruction losses to learn the token-level structural representation. Second, the AE is frozen, and the FM is trained by flow matching on normalized AE encodings of MD conformations, learning to generate latent states from Gaussian noise. Third, the FM and RM are jointly trained with ensemble-level supervision, and LoRA is applied to the trunk while the AE remains frozen.

Joint training supervises the final refined ensemble through its mean positions, fluctuation amplitudes and joint coordinate covariance:
\begin{equation}
 \mathcal L_{\mathrm{joint}}
 =\lambda_{\mu}\mathcal L_{\mathrm{mean}}
 +\lambda_v\mathcal L_{\mathrm{var}}
 +\lambda_c\mathcal L_{\mathrm{cov}}
 +\mathcal L_{\mathrm{aux}}.
 \label{eq:main_ensemble_objective}
\end{equation}
The mean objective matches average atomic positions, and the variance objective matches root-mean-square fluctuation amplitudes. The covariance objective compares the full, weighted joint coordinate covariance of selected representative atoms after trace normalization. It captures both the directional variability of individual atoms and correlated displacements between atoms. Trace normalization emphasizes the relative structure of these fluctuations, complementing the amplitude information supplied by the variance objective. These statistics are compared in a shared coordinate frame without requiring one-to-one pairing of generated and MD conformations.

The remaining terms, collected in $\mathcal L_{\mathrm{aux}}$, include flow matching and noise-conditioned coordinate denoising on real MD conformations, bidirectional ensemble coverage, geometric penalties, and preservation of proposal geometry through pair distances. Together, the objectives connect the statistical fidelity of the ensemble with the structural quality of its constituent conformations. Loss definitions, atom selections, weighting schemes and optimization settings are detailed in Appendix~\ref{sec:si_training}.

\subsection{Inference}
\label{sec:methods_inference}

Inference is performed under fixed molecular and structural conditions, with structural inputs provided by reference or predicted conformations. Independent Gaussian latent samples are transformed by integrating the FM velocity field. After inverse normalization, the AE decoder converts the sampled latent states into all-atom proposals for exploration across conformational basins.

The RM then directly generates all-atom conformations through proposal-conditioned diffusion, describing within-basin variability while improving local geometry. The resulting conformations form an ensemble without temporal ordering. Latent integration, conditional diffusion sampling and reference settings are provided in Appendix~\ref{sec:si_inference}.

%% file: sections/020evaluation.tex
\subsection{Evaluation protocols}
\label{sec:metrics_overview}
\label{sec:main_evaluation_protocols}

We evaluated conformational variability, distributional agreement, interactions and reference-state recovery using matched atoms and benchmark-specific alignments. Mean pairwise RMSD measured ensemble breadth, while atom-wise RMSF measured flexibility. For aligned coordinates $X_{ka}$, with $K$ conformations, $n$ atoms and atom-wise mean $\mu_a$, RMSF and root-mean-squared Wasserstein discrepancy (RMWD) were
\begin{equation}
\begin{aligned}
\mathrm{RMSF}_a &= \left[K^{-1}\sum_{k=1}^{K}\|X_{ka}-\mu_a\|_2^2\right]^{1/2}, \\
\mathrm{RMWD}(P,R) &= \left[n^{-1}\sum_{a=1}^{n}W_2^2\!\left(\mathcal{N}_a^P,\mathcal{N}_a^R\right)\right]^{1/2},
\end{aligned}
\end{equation}
where $\mathcal N_a^P$ and $\mathcal N_a^R$ are Gaussian approximations to predicted and reference atomic coordinates. We denote this discrepancy as Root mean $W_2$ distance (RMWD). Pairwise RMSD $r$ compares mean ensemble breadth across targets; Global RMSF $r$ concatenates matched atomic RMSF values across targets; Per-target RMSF $r$ compares profiles within individual systems and is summarized by the median for the MD benchmarks and the mean for DD-13M. Here, $r$ denotes Pearson correlation. Distributional accuracy also uses Joint PCA $W_2$ distance or MD PCA $W_2$ distance, according to the fitted basis, and scalar-observable $W_1$. Ensemble observables include Weak contacts $J$, Transient contacts $J$, Exposed residue $J$ and Exposed MI matrix $\rho$, where $J$ denotes Jaccard similarity and $\rho$ denotes Spearman correlation; contact-probability errors and amplitude ratios provide complementary summaries.

For ConforMix and ProMiSE, dual-state success required recovery of both experimental states with TM-score $>0.8$ and greater similarity to the matched than the opposite state; ConforMix additionally used an RMSD criterion of half the reference-state separation. ProMiSE responses were assessed by displacement-direction cosine, residue-displacement Spearman correlation and AP20 (average precision for the most displaced 20\% of residues). For DD-13M, weighted pose coverage was $C_\delta=\sum_j w_j\mathbf1[\min_i d(P_i,R_j)\leq\delta]$ and precision was $P_\delta=K^{-1}\sum_i\mathbf1[\min_j d(P_i,R_j)\leq\delta]$, where $w_j$ are normalized MD weights and $d$ is symmetry-corrected ligand RMSD after protein alignment; thresholds were 2 and 5~\AA{}. Comparisons used matched target sets and sampling budgets, with bootstrap intervals over targets or dependency blocks: 90\% for the ligand flexibility correlations in Fig.~\ref{fig:misato_benchmark_overview}a\textendash{}c and 95\% for other comparisons. Benchmark-specific atom selections, reference frames and statistical procedures are detailed in Appendix~\ref{sec:eval_protocols}; ablation analyses and figures are presented in Appendix~\ref{sec:si_ablation}, with additional comparisons in Supplementary Results (Appendix~\ref{sec:si_results}).

%% file: sections/040conclusion.tex
\par\addvspace{8pt}
\noindent\begin{minipage}{\linewidth}
\setlength{\parskip}{0pt}
\section*{Contributions}

\textbf{Project Lead}\par\nobreak\vspace{2pt}

Peicong Lin$^{1}$, Ziqiao Zhang$^{1}$

\par\vspace{6pt}
\textbf{Overall Technical Lead}\par\nobreak\vspace{2pt}

Shuangjia Zheng$^{2,1}$ provided overall technical supervision.

\par\vspace{6pt}
\textbf{Contributors}\par\nobreak\vspace{2pt}

Jiahao Yang$^{1,2,*}$, Dong Liu$^{1,*}$, Hongliang Li$^{1,3}$, Ji Chen$^{1}$, Yuehua Liu$^{1}$, Daoyun Ding$^{1}$, Duo An$^{1}$, Jun Li$^{1}$, Jiahua Rao$^{4}$, Ziqiao Zhang$^{1}$, Peicong Lin$^{1}$, Shuangjia Zheng$^{2,1}$

\par\nobreak\vspace{3pt}
$^{*}$Equal contribution.

\par\vspace{6pt}
\textbf{Affiliation}\par\nobreak\vspace{2pt}

$^1$Lingang Laboratory, Shanghai, China\par\vspace{1pt}

$^2$Shanghai Jiao Tong University, Shanghai, China\par\vspace{1pt}

$^3$Fudan University, Shanghai, China\par\vspace{1pt}

$^4$Sun Yat-sen University, Guangzhou, China\par

\par\vspace{10pt}
\section*{Acknowledgments}

We thank Bo Yuan, Dingyan Wang, Shuya Li, Yanfeng Chen, Xuan Liu, Leyun Wu, Runze Ma, Yuyang Tao, Yiwei Yang for their valuable discussions and insightful feedback.
\end{minipage}

%% file: sections/SI.tex
\setcounter{figure}{0}
\renewcommand{\thefigure}{\arabic{figure}}
\renewcommand{\theHfigure}{supp.\arabic{figure}}
\renewcommand{\theHequation}{supp.\arabic{equation}}
\captionsetup[figure]{name=Supplementary Figure}

\input{sections/SI_data}
\input{sections/SI_method}

\input{sections/SI_evaluation}
\input{sections/SI_ablation}
\FloatBarrier
\input{sections/SI_results}

%% file: sections/SI_data.tex
\section{Data}
\label{sec:supp_data_review}

\subsection{Molecular dynamics data}
\paragraph{ATLAS.}
ATLAS provides atomistic simulations of protein monomers. We used the designated protein chain for each target and maintained correspondence between its initialization structure and MD conformations through residue and named-atom mappings. Initialization structures were stored separately from the trajectory sampling pool.

\paragraph{MISATO.}
MISATO provides MD conformations of protein\textendash{}ligand complexes. We used the processed subset containing one protein chain and one ligand per target, with atom identities mapped from the source HDF5 representation. Protein and ligand coordinates were treated as components of a common molecular system, preserving their relative geometry.

\paragraph{mdCATH.}
mdCATH provides MD simulations of structurally diverse protein domains. The modeled target was the represented domain, with residue and named-atom mappings maintained across conformations. Longer sequences available as multiple-sequence-alignment context were not treated as additional modeled residues when comparing target identities.

\paragraph{DynaRepo-PPI.}
DynaRepo-PPI denotes the protein\textendash{}protein evaluation cohort assembled from MD trajectories in DynaRepo~\cite{dynarepo}. These trajectories describe fluctuations within multichain assemblies, providing reference ensembles for whole-complex flexibility, interface motion and interchain contacts. Predicted and reference conformations were compared using matched atoms in a common coordinate frame, preserving the relative placement of molecular partners.

\subsection{Molecular correspondence and ensemble sampling}
Each processed target retained its molecular identities, structural reference, atom correspondence, coordinate-validity masks and MD conformations. A valid simulation coordinate was distinguished from experimental observability; masks determined which atoms contributed to supervision. Joint training selected ATLAS, MISATO or mdCATH with equal probability, followed by uniform target selection within the chosen source and sampling from that target's conformations. The ensemble objectives used conformational samples and ensemble statistics without supervision on temporal ordering or transition times.

\subsection{Internal partitions and target identity}
Each dataset underwent source-specific preprocessing to remove duplicate targets and target-level redundancy. All conformations of a molecular target were assigned to the same partition. Cross-partition overlap checks used complete molecular composition, including component identities and stoichiometry, to prevent repeated targets from contributing to both model development and evaluation. Evaluation used the corresponding public test partitions or curated public benchmark cohorts.

\subsection{Functional-state and interaction benchmarks}
ConforMix provides paired experimental conformations spanning cryptic-pocket rearrangements, domain motions, fold switching and membrane transporters. ProMiSE provides matched apo\textendash{}holo structures for ligand- and protein-dependent conformational responses. Methods were compared on matched cohorts within each benchmark. The final application cohorts shared no complete molecular targets with the parent MD training corpus.

\subsection{DD-13M adaptation and evaluation}
DD-13M provides protein\textendash{}ligand dissociation trajectories, with separate partitions for model adaptation, validation and final evaluation. The source partitions were constructed using sequence components defined by a 40\% sequence-identity threshold and at least 80\% coverage on both sides of an alignment; source PDB identifiers and sequence components were kept within a single partition. Comparisons used 300 predicted conformations per system and method and up to 300 MD reference frames, retaining the available reference frames and their original weights.

\subsection{Training-frame preparation}
\label{sec:si_data}
Atom correspondence was established through molecular, residue and atom identities, with coordinates and validity masks kept in the same order. Protein atom37 representations were mapped to model atom arrays, whereas ligand atoms retained explicit chemical identities. Coordinates were expressed in \AA{}. Simulation-valid atoms could contribute to supervision even when unresolved experimentally; invalid coordinates remained masked.

AE training used a single frame per micro-step, whereas flow and joint training used multiple conformations from the sampled system. Whole-system centering and rotation augmentation acted jointly on molecular partners to preserve their relative placement. Reconstruction compared coordinates in the same augmented frame; ensemble statistics used a shared structural pose and common valid atom support. Frames supplied structural examples and empirical ensemble statistics rather than transition labels.

%% file: sections/SI_method.tex
\input{sections/SI_architecture}

\input{sections/SI_training}

\input{sections/SI_inference}

%% file: sections/SI_architecture.tex
\section{Architecture}
\label{sec:si_architecture}

OriDynamics implements latent ensemble decomposition through a deterministic autoencoder (AE), a latent flow module (FM), and an all-atom refine module (RM). The AE establishes a structural representation, the FM and decoder generate conformational proposals for global exploration across basins, and the RM performs local exploitation through proposal-conditioned sampling of within-basin variability at all-atom resolution.

\subsection{Molecular conditioning}
\label{sec:si_trunk}

The molecular trunk embeds molecular identity and chemical features while
retaining atom-to-token mappings. Its MSA module incorporates available
protein evolutionary information, and the Pairformer updates single and
pair representations through pair-biased attention and triangle operations.
For $N$ molecular tokens, the trunk produces single representations
$s\in\mathbb R^{N\times384}$ and pair representations
$z\in\mathbb R^{N\times N\times128}$, together with atom-level features.

We use $C$ to denote the molecular inputs and supplied structural information
held fixed when sampling an ensemble. The AE uses fixed trunk context.
During joint training, LoRA adapts the MSA and Pairformer context supplied to
the FM and RM, while the AE context remains unchanged. This preserves the
learned encoding and decoding convention as ensemble generation is adapted.

\subsection{Structural representation and decoding}
\label{sec:si_codec}

The AE combines atom attention, token-level structural processing, and
all-atom decoding. For a conformation $X\in\mathbb R^{A\times3}$ containing
$A$ atoms, the encoder produces one 32-dimensional latent vector per token:
\begin{equation}
 h=\mathcal E_{\vartheta}(X;C)\in\mathbb R^{N\times32},\qquad
 \widehat X=\mathcal D_{\psi}(h;C)\in\mathbb R^{A\times3}.
 \label{eq:si_codec}
\end{equation}
Here, $\mathcal D_{\psi}$ denotes the complete coordinate reconstruction
map. The AE is initialized from a pretrained coordinate diffusion
architecture and trained deterministically, without posterior sampling or
KL regularization. Reconstruction establishes a latent space of
conformational variation while preserving molecular identities and atom
correspondence. After pretraining, the same frozen decoder converts
FM-generated latent states into all-atom proposals.

\subsection{Proposal generation and conditional refinement}
\label{sec:si_flow_architecture}

The FM learns a conditional velocity field in the normalized AE latent
space. A transformer processes the latent tokens, with the flow
interpolation variable and single features providing adaptive modulation,
and pair features supplying attention biases. Integrating this velocity
field transforms Gaussian latent noise into conformational representations.
Inverse normalization and decoding produce all-atom proposals
$X_{\mathrm{prop}}=\mathcal D_{\psi}(\mathcal F_{\phi}(\epsilon_z;C);C)$,
where $\epsilon_z\sim\mathcal N(0,I)$ and $\mathcal F_{\phi}$ includes
inverse normalization.

The RM uses a separately parameterized all-atom diffusion network adapted
from an AlphaFold 3\textendash{}like module. Its denoiser predicts clean all-atom
coordinates from a noisy coordinate state, the noise level, and molecular
context. The decoded proposal conditions the complete diffusion generation
process, which we write as
$X=\mathcal R_{\omega}(X_{\mathrm{prop}};C)$.
The RM directly generates all-atom conformations conditioned on the
proposal, describing variability within its basin while improving
molecular geometry.

The two generative modules compose a distribution over final conformations:
\begin{equation}
 P(\,\cdot\mid C)=\int
 P_{\mathrm{refine}}(\,\cdot\mid X_{\mathrm{prop}},C)
 \,\mathrm dP_{\mathrm{prop}}(X_{\mathrm{prop}}\mid C).
 \label{eq:si_decomposition}
\end{equation}
Here $P_{\mathrm{prop}}$ is induced by latent sampling and decoding, and
$P_{\mathrm{refine}}$ by proposal-conditioned diffusion generation.
Viewing the final coordinates $X$ as a vector, the law of total covariance
gives
\begin{equation}
\begin{aligned}
 \operatorname{Cov}(X\mid C)
 ={}&\operatorname{Cov}\!\left(
       \mathbb E[X\mid X_{\mathrm{prop}},C]\mid C\right)\\
 &+\mathbb E\!\left[
       \operatorname{Cov}(X\mid X_{\mathrm{prop}},C)\mid C\right].
\end{aligned}
 \label{eq:si_total_covariance}
\end{equation}
The first term describes variation among conditional mean conformations,
and the second describes average variation within conditional ensembles.
A proposal supplies the condition and need not equal the conditional mean.
Ensemble supervision constrains their combined contribution through final
samples, without requiring explicit basin labels or separate covariance
estimates for each proposal-conditioned ensemble
(Appendix~\ref{sec:si_losses}).

%% file: sections/SI_training.tex
\section{Training}
\label{sec:si_training}
\label{sec:si_methods}

\subsection{Training the structural representation and generator}

\subsubsection{Autoencoder pretraining}
\label{sec:si_ae_training}
The AE learns token-level conformational representations by reconstructing
individual MD conformations under fixed trunk context. Its objective
combines coordinate MSE and pair-distance reconstruction:
\begin{equation}
 \mathcal L_{\mathrm{AE}}
 =\mathcal L_{\mathrm{coord}}
 +\lambda_{\mathrm{pair}}\mathcal L_{\mathrm{pair}}.
 \label{eq:si_ae_loss}
\end{equation}
Both terms use valid atom support. Coordinate errors are balanced across
molecular entities, and pair-distance errors across entity-pair groups.
Reconstruction is evaluated in the shared target frame without an
independent rigid fit.

\subsubsection{Flow-matching training}
\label{sec:si_flow}
The pretrained AE is frozen, and the FM learns the distribution of its
encodings of MD conformations. Channelwise means $\mu_h$ and standard
deviations $\sigma_h$, estimated from training data only, define normalized
latents $u=(h-\mu_h)/\sigma_h$. Flow matching uses
\begin{equation}
\begin{aligned}
 u_t&=(1-t)\epsilon+tu,\\
 \mathcal L_{\mathrm{FM}}
 &=\mathbb E\!\left[
 \|v_{\phi}(u_t,t\mid C)-(u-\epsilon)\|_{\mathrm{masked}}^2
 \right],
\end{aligned}
 \label{eq:si_flow_matching}
\end{equation}
where $\epsilon\sim\mathcal N(0,I)$ and $t\sim\mathcal U(0,1)$.
The squared error is averaged over valid token\textendash{}channel entries. Only
the FM is updated at this stage. The interpolation variable parameterizes
a path from noise to latent data, rather than physical time.

\subsubsection{Joint ensemble training}
\label{sec:si_joint}
The final stage jointly updates the FM, RM, and LoRA parameters in the MSA
and Pairformer modules. The AE and its trunk context remain fixed.
Latent sampling and decoding produce conformational proposals, and the RM
generates all-atom conformations conditioned on these proposals. Joint
training combines ensemble-level and conformation-level supervision:
\begin{equation}
\begin{aligned}
 \mathcal L_{\mathrm{joint}}
 &=\mathcal L_{\mathrm{ensemble}}+\mathcal L_{\mathrm{sample}},\\
 \mathcal L_{\mathrm{ensemble}}
 &=\lambda_\mu\mathcal L_{\mathrm{mean}}
   +\lambda_v\mathcal L_{\mathrm{var}}
   +\lambda_c\mathcal L_{\mathrm{cov}}
   +\lambda_{\mathrm{cover}}\mathcal L_{\mathrm{cover}},\\
 \mathcal L_{\mathrm{sample}}
 &=\lambda_{\mathrm{replay}}\mathcal L_{\mathrm{replay}}
   +\lambda_b\mathcal L_{\mathrm{bond}}
   +\lambda_{\mathrm{clash}}\mathcal L_{\mathrm{clash}}\\
 &\quad+\lambda_p\mathcal L_{\mathrm{preserve}}
   +\lambda_{\mathrm{FM}}\mathcal L_{\mathrm{FM}}.
\end{aligned}
 \label{eq:si_joint_groups}
\end{equation}
Ensemble-level terms compare collections of conformations, while conformation-level terms supervise individual structures and their latent representations. In the main-text objective (Eq.~\ref{eq:main_ensemble_objective}), $\mathcal L_{\mathrm{aux}}$ collects the weighted coverage term $\lambda_{\mathrm{cover}}\mathcal L_{\mathrm{cover}}$ and the conformation-level objectives $\mathcal L_{\mathrm{sample}}$.

\subsection{Why ensemble statistics supplement pointwise supervision}
\label{sec:si_mse_motivation}

At fixed molecular condition $C$, let $Z$ and $Y$ be independently drawn
generated and reference coordinate vectors in a common frame with
consistent atom correspondence. Assume finite second moments, and write
their conditional means as $\mu_P,\mu_T$
and conditional covariances as $\Sigma_P,\Sigma_T$. Expanding the squared
error and using independence gives
\begin{equation}
 \mathbb E[\|Z-Y\|_2^2\mid C]
 =\|\mu_P-\mu_T\|_2^2
  +\operatorname{tr}\Sigma_P+\operatorname{tr}\Sigma_T.
 \label{eq:si_unpaired_mse}
\end{equation}
Because the reference term is constant, unconstrained minimization over
the generated distribution is attained at $\mu_P=\mu_T$ and $\Sigma_P=0$.
Thus, arbitrarily pairing generated samples with MD frames can reward
collapse to the mean, which may lie between populated conformational
states. This argument concerns independently paired ensemble samples,
with no other loss or geometric constraint. AE reconstruction and replay denoising
instead use paired targets: an encoded or noise-corrupted real conformation
is mapped back to that same conformation. These objectives do not satisfy
the independence assumption above and remain useful during training.

Matching means and marginal fluctuation amplitudes also leaves joint
variation undetermined. For example, zero-mean two-coordinate distributions
with
\begin{equation}
 \Sigma_{\pm}=
 \begin{pmatrix}1&\pm\rho\\ \pm\rho&1\end{pmatrix},
 \qquad 0<\rho<1,
 \label{eq:si_covariance_counterexample}
\end{equation}
have identical marginal variances but opposite displacement correlations.
Under equal coordinate weights, their trace-normalized covariance loss
is $2\rho^2>0$, although the mean and fluctuation-amplitude losses vanish.
This motivates separate constraints on mean positions, fluctuation
amplitudes, and joint covariance. These moments describe complementary
properties; they do not uniquely determine a multimodal distribution,
which is also learned through flow matching and constrained by coverage.

\subsection{Ensemble-level objectives}
\label{sec:si_losses}

Let $P\in\mathbb R^{K\times n\times3}$ and
$T\in\mathbb R^{M\times n\times3}$ denote generated and reference ensembles
on $n$ common valid representative atoms. These are protein C$\alpha$
atoms and, for PL systems, ligand heavy atoms. All ensemble statistics use
a shared coordinate frame, with interacting partners transformed together.
Generated conformations are not independently fitted within these
objectives, and no one-to-one sample assignment is required. Statistics
are computed over the pooled final RM-generated conformations.

\paragraph{Mean positions and fluctuation amplitudes.}
With $\bar P_a=K^{-1}\sum_kP_{ka}$,
$\bar T_a=M^{-1}\sum_mT_{ma}$, atom weights $w_a>0$, and $W=\sum_aw_a$,
\begin{align}
 \mathcal L_{\mathrm{mean}}
 &=\frac{1}{3W}\sum_aw_a\|\bar P_a-\bar T_a\|_2^2,\\
 V_a^P&=\frac1K\sum_k\|P_{ka}-\bar P_a\|_2^2,\qquad
 V_a^T=\frac1M\sum_m\|T_{ma}-\bar T_a\|_2^2,\\
 \mathcal L_{\mathrm{var}}
 &=\frac1W\sum_aw_a
 \left(\sqrt{V_a^P+\varepsilon_v}
              -\sqrt{V_a^T+\varepsilon_v}\right)^2.
 \label{eq:si_moments}
\end{align}
The positive $\varepsilon_v$ stabilizes the square roots.
$\mathcal L_{\mathrm{var}}$ therefore compares RMSF amplitudes.
Protein-only systems use uniform weights; PL systems assign equal total
weight to protein C$\alpha$ atoms and ligand heavy atoms.

\paragraph{Joint coordinate covariance.}
For flattened coordinates $y_k^P\in\mathbb R^{3n}$, define
\begin{equation}
\begin{aligned}
 \Sigma^P&=\frac1K\sum_k
 (y_k^P-\bar y^P)(y_k^P-\bar y^P)^\top,\\
 D_w&=\operatorname{diag}(\sqrt{w_1}I_3,\ldots,\sqrt{w_n}I_3),
\end{aligned}
 \label{eq:si_weighted_covariance}
\end{equation}
with $\Sigma^T$ defined analogously. Away from the numerical floor,
the weighted, trace-normalized covariances are
\begin{equation}
 \widehat\Sigma^P=
 \frac{D_w\Sigma^PD_w}{\operatorname{tr}(D_w\Sigma^PD_w)},\qquad
 \widehat\Sigma^T=
 \frac{D_w\Sigma^TD_w}{\operatorname{tr}(D_w\Sigma^TD_w)}.
 \label{eq:si_covariance_normalization}
\end{equation}
The objective is
\begin{equation}
 \mathcal L_{\mathrm{cov}}
 =\|\widehat\Sigma^P-\widehat\Sigma^T\|_F^2.
 \label{eq:si_covariance}
\end{equation}
Diagonal $3\times3$ blocks describe single-atom directional variability,
and off-diagonal blocks describe correlated displacements between atoms.
Trace normalization emphasizes the relative structure of joint
fluctuations, complementing the absolute amplitudes supplied by
$\mathcal L_{\mathrm{var}}$.

\paragraph{Conformational coverage.}
Bidirectional nearest-neighbor coverage supplements the moment objectives:
\begin{equation}
 \mathcal L_{\mathrm{cover}}=
 \frac12\left[
 \frac1K\sum_k\min_m d_w(P_k,T_m)^2+
 \frac1M\sum_m\min_k d_w(P_k,T_m)^2
 \right],
 \label{eq:si_coverage}
\end{equation}
where $d_w(X,Y)^2=W^{-1}\sum_aw_a\|X_a-Y_a\|_2^2$ on the
representative atoms. The two directions encourage generated samples to
remain close to reference conformations and reference conformations to
be represented by the generated ensemble.

\subsection{Conformation-level objectives and replay}
\label{sec:si_auxiliary_training}

Replay retains supervision on real MD conformations during joint training.
It includes noise-conditioned coordinate MSE, $\mathcal L_{\mathrm{MSE}}$,
and local-distance supervision, $\mathcal L_{\mathrm{lddt}}$, together with
structural auxiliary terms. The coordinate objective reconstructs a clean
MD conformation from its noise-corrupted input, and the local-distance
objective encourages agreement in local geometry. Replay uses paired
conformations and their validity masks; these losses complement the
unpaired ensemble statistics defined above.

On generated conformations, bond-length penalties use reference-ensemble
mean bond lengths, clash penalties discourage steric overlap, and
pair-distance preservation relates each refined conformation to its
decoded proposal. These constraints support molecular geometry while
allowing conformational variation. Flow matching on encoded real MD
frames remains active, retaining supervision of the latent velocity
field alongside the final-coordinate objectives.

\begin{table}[htbp]
\centering
\small
\setlength{\tabcolsep}{5pt}
\renewcommand{\arraystretch}{1.12}
\caption{Ensemble-level and conformation-level supervision during joint
training. Ensemble objectives act on the pooled final conformations.}
\label{tab:si_objectives}
\begin{tabular}{@{}p{0.19\linewidth}p{0.33\linewidth}p{0.39\linewidth}@{}}
\toprule
Level & Objectives & Role \\
\midrule
Ensemble & Mean, RMSF, joint covariance & Average positions, fluctuation amplitudes, and coordinated displacements \\
Ensemble & Bidirectional coverage & Agreement in conformational support \\
Conformation & Replay & Coordinate MSE, local-distance, and structural supervision on real MD \\
Conformation & Bond, clash, proposal preservation & Structural consistency of generated conformations \\
Conformation & Flow matching & Per-conformation latent velocity supervision \\
\bottomrule
\end{tabular}
\end{table}

\subsection{Optimization}
\label{sec:si_optimization}

Training uses AdamW with weight decay $10^{-2}$, gradient clipping, and
mixed precision. AE pretraining and FM pretraining each use a learning
rate of $10^{-4}$. FM pretraining maintains an exponential moving average
of the parameters with decay $0.999$. Joint fine-tuning of the shared model
uses separate optimizer groups for the FM, RM, and trunk LoRA parameters; all three
groups use the same peak learning rate of $3\times10^{-5}$. A linear
warm-up over 200 optimizer steps precedes the constant-rate portion
used for the reported joint checkpoint. The AE and its context remain
frozen throughout FM pretraining and joint fine-tuning.

%% file: sections/SI_inference.tex
\section{Inference}
\label{sec:si_inference}

We follow the ensemble-emulation setting used by AlphaFlow~\cite{alphaflow},
generating independent conformations for a fixed molecular system.
OriDynamics implements this setting through latent proposal generation and
proposal-conditioned all-atom diffusion.

\subsection{Conformational proposal generation}
\label{sec:si_representative_sampling}

For a fixed molecular specification and supplied structural condition $C$,
sampling starts from $u_0\sim\mathcal N(0,I)$ in the normalized latent
space. Heun integration of $\mathrm d u_t/\mathrm dt=v_\phi(u_t,t\mid C)$
from $t=0$ to $t=1$ produces the terminal state $u_1$, which is returned to
the AE latent scale and decoded:
\begin{equation}
 h=\mu_h+\sigma_h\odot u_1,\qquad
 X_{\mathrm{prop}}=\mathcal D_{\psi}(h;C).
 \label{eq:si_representative_decoding}
\end{equation}
The normalization statistics are those estimated during training, and
$\odot$ denotes elementwise multiplication. Independent latent draws
produce conformational proposals under the same input conditions.

\subsection{Conditional all-atom diffusion}
\label{sec:si_refinement_sampling}

The RM uses the decoded proposal as a structural condition for all-atom
diffusion generation. Its denoiser predicts clean coordinates at successive
noise levels, and coordinate updates follow a descending diffusion-noise
schedule. The complete generation process is written as
\begin{equation}
 X=\mathcal R_{\omega}(X_{\mathrm{prop}};C).
 \label{eq:si_conditional_sampling}
\end{equation}
This process describes local conformational variability while improving
molecular geometry. The final ensemble collects the generated conformations
across proposals, retaining atom identities and validity masks. Neither
latent integration nor coordinate-denoising steps correspond to physical
time, and the generated ensemble is unordered.

\subsection{Structural references and model use}
\label{sec:si_inference_settings}

Structural conditioning can use a supplied reference structure or a
predicted structure. The structural-reference comparison changes this
reference without changing model parameters. The AE+FM ablation retains
decoded proposals before RM generation; both comparisons are described in
Appendix~\ref{sec:si_ablation}. The four main benchmarks, ConforMix, and ProMiSE use the same jointly trained model. DD-13M uses a checkpoint adapted on its training systems and selected using validation data.

%% file: sections/SI_evaluation.tex
\section{Evaluation}
\label{sec:eval_protocols}
\label{sec:eval_common}
\label{sec:eval_monomer}
\label{sec:methods_monomer_evaluation}
\label{sec:eval_misato}
\label{sec:eval_dynappi}
\label{sec:eval_conformix}
\label{sec:eval_promise}
\label{sec:eval_dd13m}

\paragraph{Benchmark and baseline sources.}
MD ensemble evaluation uses ATLAS~\cite{atlas}, mdCATH~\cite{mdcath}, MISATO~\cite{misato} and the DynaRepo-derived protein\textendash{}protein cohort DynaRepo-PPI~\cite{dynarepo}; the application benchmarks are ConforMix~\cite{oridata_conformix2025}, ProMiSE~\cite{oridata_promise2026} and DD-13M~\cite{oridata_dd13m2026}. Comparators include AlphaFlow/AlphaFlow-MD and ESMFlow~\cite{alphaflow}, MDGen~\cite{mdgen}, BioEmu~\cite{bioemu}, BioEmu2~\cite{bioemu2}, BioKinema~\cite{biokinema}, Boltz-2~\cite{boltz2}, Protenix-v1/v2~\cite{protenix}, ESMFold2~\cite{esmfold2} and UnbindingFlow~\cite{oridata_dd13m2026}. Versioned papers and official model repositories are provided in the references; the model-generation labels in each figure distinguish comparator families. The frozen ProMiSE and DD-13M source snapshots were \texttt{bfcaac9} and \texttt{b7fdae5}, respectively.

\paragraph{Atom correspondence and alignment.}
Predicted ($P$) and reference ($R$) ensembles are aligned on matched atoms in a shared coordinate frame. Each complex undergoes a single rigid transformation, preserving the relative placement of its molecular partners. Monomer metrics use matched C$\alpha$ atoms aligned to the initial reference structure. MISATO ligand and pocket metrics use heavy atoms after full-complex alignment; the pocket includes the ligand and protein atoms within 10~\AA{} of it in MD frame~0. On DynaRepo-PPI, RMSF correlations use heavy atoms, whereas pairwise RMSD, RMSF amplitude errors, contacts and PCA use C$\alpha$ atoms; interfaces are defined by the fixed 10~\AA{} reference mask. Reference ensembles contain 301 ATLAS, 400 mdCATH and 100 complex-benchmark MD frames per target. DD-13M retains normalized MD frame weights. Coordinate-based distances are reported in \AA{}.

\paragraph{Conformational breadth and flexibility.}
For aligned coordinates $X_{ka}$, with $K$ conformations and $n$ evaluated atoms, let $\mu_a=K^{-1}\sum_kX_{ka}$. Pairwise RMSD and atom-wise RMSF are
\begin{align}
 d(X_k,X_l)&=\left[\frac1n\sum_{a=1}^{n}\|X_{ka}-X_{la}\|_2^2\right]^{1/2},
 \label{eq:eval_pairwise_rmsd}\\
 \mathrm{RMSF}_a(X)&=\left[\frac1K\sum_{k=1}^{K}\|X_{ka}-\mu_a\|_2^2\right]^{1/2}.
 \label{eq:eval_rmsf}
\end{align}
Mean pairwise RMSD averages distances in this shared frame without additional pairwise fitting. The monomer implementation includes all prediction pairs, including the zero diagonal, and uses two fixed reference-frame draws. Pairwise RMSD $r$ is the Pearson correlation of predicted and reference mean breadth across targets. Global RMSF $r$ concatenates matched atomic RMSF values across targets before computing Pearson correlation; Per-target RMSF $r$ computes this correlation separately within each target, then reports the median for the MD benchmarks and the mean for DD-13M. Correlation measures profile agreement, while amplitude ratios assess fluctuation magnitude; a prediction/MD ratio of one denotes equal amplitude. Weighted reference means and RMSF replace uniform frame averages by normalized weights.

\paragraph{Distributional agreement.}
Root mean $W_2$ distance (RMWD) compares Gaussian approximations to marginal atomic coordinate distributions~\cite{alphaflow}. With atomic means $\mu_a$ and population covariances $\Sigma_a$, the squared Gaussian $W_2$ distance and its aggregate are
\begin{align}
 g_a^2&=\|\mu_a^P-\mu_a^R\|_2^2
 +\operatorname{tr}\!\left(\Sigma_a^P+\Sigma_a^R\right)\notag\\
 &\quad-2\operatorname{tr}\!\left[\left((\Sigma_a^R)^{1/2}\Sigma_a^P(\Sigma_a^R)^{1/2}\right)^{1/2}\right],
 \label{eq:eval_atomic_w2}\\
 \mathrm{RMWD}(P,R)&=\left(\frac1n\sum_{a=1}^{n}g_a^2\right)^{1/2}.
 \label{eq:eval_rmwd}
\end{align}
The mean-position and covariance contributions distinguish shifts in average structure from differences in fluctuations. Each $\Sigma_a=K^{-1}\sum_k(X_{ka}-\mu_a)(X_{ka}-\mu_a)^\top$ is a $3\times3$ population covariance; the training objective additionally constrains joint covariance across atoms.

PCA-based distances assess collective conformational distributions in a common basis. Joint PCA is fitted to reference and predicted coordinates together; MD PCA is fitted to reference coordinates only. For projected coordinates $u_i,v_j$ and normalized weights $p_i,q_j$,
\begin{align}
 W_{2,\mathrm{PCA}}(P,R)&=\left[\min_{\pi\in\Pi(p,q)}\sum_{i,j}\pi_{ij}\frac{\|u_i-v_j\|_2^2}{n}\right]^{1/2},
 \label{eq:eval_pca_w2}\\
 \Pi(p,q)&=\{\pi\geq0:\pi\mathbf1=p,\ \pi^\top\mathbf1=q\}.\notag
\end{align}
The monomer and DynaRepo-PPI metrics use the first two components and minimum-cost assignment between equally weighted, equal-sized samples. DD-13M uses weighted MD PCA and balanced optimal transport; its coordinates are normalized by $\sqrt n$ before projection, so atom-count normalization is applied only once. For scalar observables, including interaction distances, native-contact fractions and centroid displacements, $W_1(P,R)=\int|F_P(t)-F_R(t)|\,\mathrm dt$, where $F_P,F_R$ are empirical cumulative distributions. Lower Wasserstein distances indicate closer agreement.

\paragraph{Contacts and solvent exposure.}
Contact probability is the ensemble-weighted frequency of a distance falling below its threshold. DynaRepo-PPI contacts use an 8~\AA{} C$\alpha$ cutoff. Weak contacts are present in the input structure but have ensemble probability below 0.9; transient contacts are absent from the input but have probability above 0.1. Weak contacts $J$, Transient contacts $J$ and Exposed residue $J$ compare predicted and reference sets using $J(A,B)=|A\cap B|/|A\cup B|$, with $J=0$ for an empty union in the DynaRepo-PPI implementation. Its exposed-residue sets contain residues buried in the input that become exposed in more than 10\% of frames, using a side-chain solvent-accessible surface-area threshold of $0.02~\mathrm{nm}^2$ and a 0.28~nm probe radius. Contact-probability MAE averages $|p_c^P-p_c^R|$ over a fixed set of contacts. Binary side-chain solvent-exposure states define pairwise mutual information,
\begin{equation}
 I(E_a;E_b)=\sum_{u,v\in\{0,1\}}p_{ab}(u,v)\log\frac{p_{ab}(u,v)}{p_a(u)p_b(v)},
 \label{eq:eval_exposure_mi}
\end{equation}
where $p_{ab}$ is the joint probability and $p_a,p_b$ are marginal probabilities; zero-probability terms are omitted. Natural logarithms give units of nats. Exposed MI matrix $\rho$ is the Spearman correlation between corresponding predicted and reference matrix entries, measuring agreement in coordinated exposure.

\paragraph{Experimental states and partner-dependent responses.}
Dual-state success requires a qualifying sample for each experimental state; different samples may recover the two states. TM-score success requires a score above 0.8 and greater similarity to the matched than the opposite state. The ConforMix RMSD criterion requires both state-specific minima to be no larger than half the reference-state separation. BioEmu2 ConforMix predictions use its official online MSA pipeline and default sampling settings. ProMiSE responses compare mean with-partner and without-partner conformations after apo alignment. Displacement-direction cosine measures directional agreement with the experimental change; residue-displacement Spearman correlation compares displacement-magnitude rankings. AP20 is average precision when residues are ranked by predicted displacement and the experimentally most displaced 20\% are treated as positives. \label{sec:eval_promise_groups}
Following ProMiSE's cluster-averaged evaluation~\cite{oridata_promise2026}, we compute $S_{\mathrm{cluster}}=G^{-1}\sum_{g=1}^{G}n_g^{-1}\sum_{i\in g}s_i$, where $s_i$ indicates dual-state success and $n_g$ is the number of pairs in a cluster. Clusters retain the benchmark's task-specific protein-group identifiers after its sequence-redundancy filtering; the original curation groups sequences at 95\% identity and removes redundancy between representatives at 40\% identity. The evaluated cohort spans 160 task\textendash{}cluster groups (85 ligand-induced and 75 protein-induced). Equal cluster weighting limits the influence of groups containing many apo\textendash{}holo pairs; equal-pair results and cluster-removal sensitivity analyses are reported alongside it. Low, Mid and High displacement refer to the experimentally determined apo\textendash{}holo C$\alpha$ RMSD $D$: $D\leq4.1925$~\AA{}, $4.1925<D\leq9.1782$~\AA{} and $D>9.1782$~\AA{}, respectively. These reference-derived boundaries were fixed before model-response comparisons and retained after overlap exclusions.

\paragraph{Ligand pose coverage and precision.}
For $K$ predictions $P_i$, reference poses $R_j$ and normalized MD weights $w_j$, coverage and precision at threshold $\delta$ are
\begin{align}
 \mathrm{Coverage}_{\delta}&=\sum_j w_j\,\mathbf1\!\left[\min_i d(P_i,R_j)\leq\delta\right],
 \label{eq:eval_coverage}\\
 \mathrm{Precision}_{\delta}&=\frac1K\sum_i\mathbf1\!\left[\min_j d(P_i,R_j)\leq\delta\right].
 \label{eq:eval_precision}
\end{align}
Here $d$ is symmetry-corrected ligand heavy-atom RMSD in the protein-aligned frame, evaluated at 2 and 5~\AA{}. Coverage measures the reference probability mass reached by predictions; precision measures the fraction of predictions close to at least one reference pose. State-specific coverage restricts reference poses to each MD-defined state and renormalizes their weights. Higher values indicate greater coverage or precision, respectively.

\paragraph{DD-13M state definitions.}
\label{sec:eval_dd13_states}
The states summarize contact loss and outward ligand displacement relative to the initialization structure. Initial protein\textendash{}ligand heavy-atom pairs within 4.5\,\text{\AA} define a contact set $\mathcal C_0$. Their smooth contact fraction is $Q(X)=|\mathcal C_0|^{-1}\sum_{(i,j)\in\mathcal C_0}[1+\exp((d_{ij}(X)-4.5\,\text{\AA})/(0.5\,\text{\AA}))]^{-1}$. We use normalized contact loss $a=\operatorname{clip}_{[0,1]}[(Q(X_0)-Q(X))/(Q(X_0)+10^{-6})]$ and normalized outward displacement $r_p=\operatorname{clip}_{[0,3]}[\max(0,d_p(X)-d_p(X_0))/(5\,\text{\AA})]$. Here $d_p$ is the distance between the ligand center of mass and the geometric center of pocket C$\alpha$ atoms; pocket residues have a heavy atom within 6~\AA{} of the initial ligand. The mutually exclusive assignments are
\begin{equation}
\begin{aligned}
S_0 &: a<0.2\ \text{and}\ r_p<0.25,\\
S_1 &: a<0.6\ \text{and}\ r_p<0.5,\ \text{excluding }S_0,\\
S_3 &: a\geq0.8\ \text{and}\ r_p\geq0.5,\\
S_2 &: \text{all remaining configurations}.
\end{aligned}
\label{eq:dd13_states}
\end{equation}
These labels describe MD reference configurations, from retained initial contacts to substantial contact loss and displacement. A target without reference support for a state is omitted from that state's summary rather than assigned zero coverage.

\paragraph{Aggregation and uncertainty.}
Comparisons retain matched target sets and sampling budgets. Ligand flexibility correlations in Fig.~\ref{fig:misato_benchmark_overview}a\textendash{}c use 90\% percentile intervals, defined by the 5th and 95th percentiles of 2,000 paired target-bootstrap estimates (seed 137). Other bootstrap confidence intervals use the 2.5th and 97.5th percentiles to give 95\% intervals. Matched targets or dependency blocks are resampled together across methods; Global RMSF $r$ resamples complete target profiles rather than individual atoms. ProMiSE and DD-13M retain their benchmark-defined dependency blocks. Monomer stratification uses MD-derived secondary-structure groups and the length bins $<100$, 100\textendash{}199 and $\geq200$ residues; $\dagger$ marks groups with fewer than five targets.

%% file: sections/SI_ablation.tex
\FloatBarrier
\section{Ablations}
\label{sec:si_ablation}

We examine conditional refinement and structural-reference source using the same jointly trained OriDynamics model across ATLAS, mdCATH, MISATO and DynaRepo-PPI. Pairwise RMSD correlation assesses agreement in how ensemble breadth varies across targets. Global RMSF correlation compares pooled atomic fluctuations, while per-target RMSF correlation assesses flexibility profiles within individual targets. Metric definitions follow Appendix~\ref{sec:eval_protocols}; MISATO scores use ligand atoms, and DynaRepo-PPI scores describe whole complexes.

\subsection{Contribution of conditional refinement}

We compare decoded proposals from latent flow sampling (AE+FM) with the final ensemble after conditional all-atom refinement (AE+FM+RM; Appendix~\ref{sec:si_inference}). Both branches use Protenix (PTX)-predicted structural inputs and the same learned AE and FM parameters. This comparison assesses the contribution of applying the RM at inference within the jointly trained model, rather than the performance of an independently trained model without refinement.

\begin{figure}[!htbp]
    \centering
    \includegraphics[width=\linewidth]{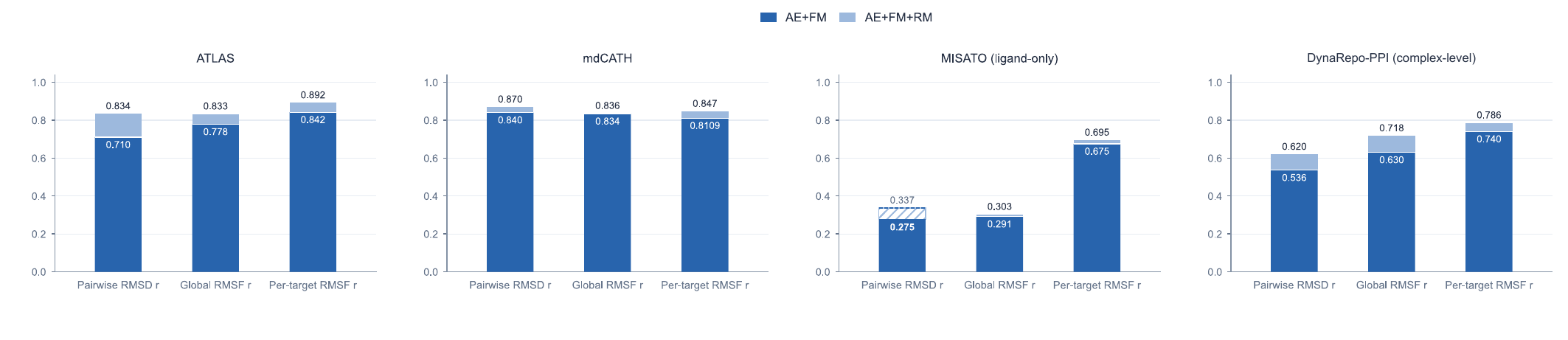}
    \caption{\textbf{Contribution of conditional refinement.}
    Panels show ATLAS, mdCATH, MISATO (ligand-only) and DynaRepo-PPI (complex-level).
    AE+FM and AE+FM+RM denote decoded proposals and refined conformations from the same trained model using PTX-predicted structures.
    For increases, dark blue reaches the AE+FM value and light blue extends to the AE+FM+RM value. For decreases, the filled bar ends at the AE+FM+RM value and the hatched segment extends to the AE+FM value.
    Labels give absolute correlations; hatched segments mark reductions and do not contribute to the final filled-bar height.}
    \label{fig:si_rm_ablation}
\end{figure}

Refinement increases 11 of the 12 reported dataset\textendash{}metric correlations, including all three metrics on ATLAS, mdCATH and DynaRepo-PPI (Supplementary Fig.~\ref{fig:si_rm_ablation}). Both global and per-target RMSF correlations improve on every dataset. These gains show that conditional all-atom generation improves agreement with reference fluctuation patterns beyond latent sampling and decoding alone, both for pooled atoms and for flexibility profiles within individual targets.

The exception is MISATO ligand pairwise RMSD correlation, which decreases from 0.337 to 0.275 despite gains in both RMSF correlations. Improved agreement in atom-wise fluctuation patterns therefore does not necessarily improve agreement in which complexes exhibit broader ligand ensembles. Conditional refinement changes these distinct aspects of the sampled distribution, and their responses need not coincide. Evaluating both ensemble breadth and atomic flexibility captures this difference.

\FloatBarrier

\subsection{Effect of the structural-reference source}

We compare a PTX-predicted structure with the supplied MD initialization structure (frame 0) as the structural reference for OriDynamics and BioEmu2. The figure labels \textit{predicted} and \textit{MD frame 0} denote these two reference sources, respectively. OriDynamics uses a structural reference in both settings with model parameters held fixed; its predicted-reference results are the AE+FM+RM results above.

\begin{figure}[!htbp]
    \centering
    \includegraphics[width=\linewidth]{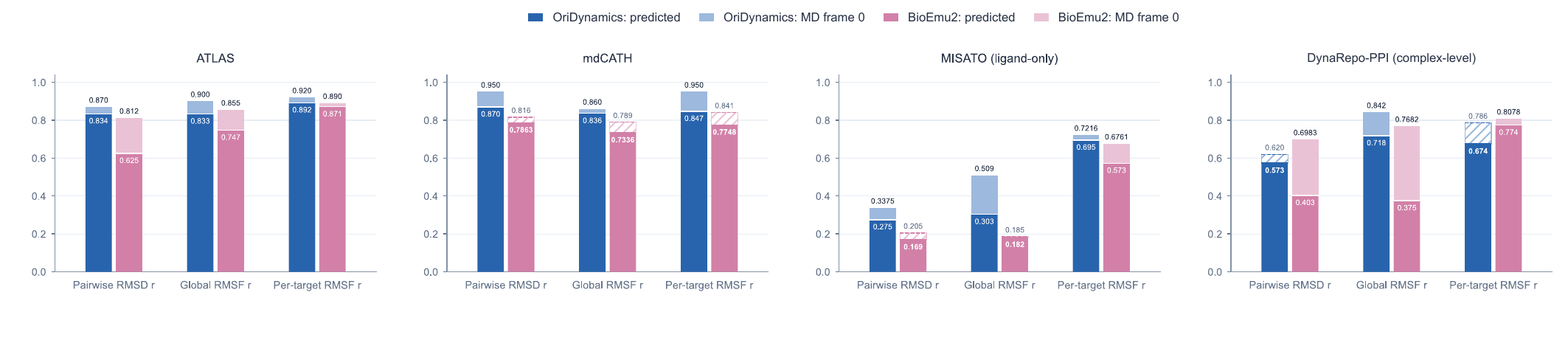}
    \caption{\textbf{Effect of structural-reference source.}
    Panels show ATLAS, mdCATH, MISATO (ligand-only) and DynaRepo-PPI (complex-level).
    Blue and pink bars denote OriDynamics and BioEmu2, respectively. The labels \textit{predicted} and \textit{MD frame 0} refer to the predicted structure and MD initialization structure, respectively.
    For increases, the saturated segment reaches the predicted-reference value and the lighter segment extends to the supplied-reference value. For decreases, the filled bar ends at the supplied-reference value and the hatched segment extends to the predicted-reference value.
    Labels give absolute correlations; hatched segments mark reductions and do not contribute to the final filled-bar height.}
    \label{fig:si_anchor_ablation}
\end{figure}

The supplied reference improves all three OriDynamics correlations on ATLAS, mdCATH and MISATO (Supplementary Fig.~\ref{fig:si_anchor_ablation}). Its geometry is consistent with the initialization of the reference simulation, which may help the generator match that simulation's structural context. A single starting conformation nevertheless does not specify the full conformational distribution.

DynaRepo-PPI shows a mixed response: global RMSF correlation improves, while pairwise RMSD and per-target RMSF correlations decrease. The Global RMSF correlation reflects variation across systems as well as across atoms, whereas the per-target correlation focuses on the flexibility profile within each complex. Their different responses show that better global agreement can coexist with lower agreement in within-complex fluctuation profiles and ensemble breadth across targets.

BioEmu2 also exhibits dataset-dependent responses: all three correlations improve on ATLAS and DynaRepo-PPI, all decline on mdCATH, and only per-target RMSF improves on MISATO. The benefit of the supplied reference therefore depends on both the model and the dataset. Structural context can assist ensemble generation, but it does not uniformly improve all ensemble observables.

\FloatBarrier

%% file: sections/SI_results.tex
\section{Supplementary Results}
\label{sec:si_results}

The figures below provide additional protein\textendash{}protein interface and functional-state comparisons, together with structural views of proposal-conditioned ensemble generation. Evaluation procedures are described in Appendix~\ref{sec:eval_protocols}.

\input{results/figure_dynappi_interface_supp}
\FloatBarrier
\input{application/figure_functional_states_supp}
\clearpage
\input{application/figure_fm_rm_landscapes_supp}

%% file: results/figure_dynappi_interface_supp.tex
\begin{figure}[H]
    \centering
    \includegraphics[width=\linewidth]{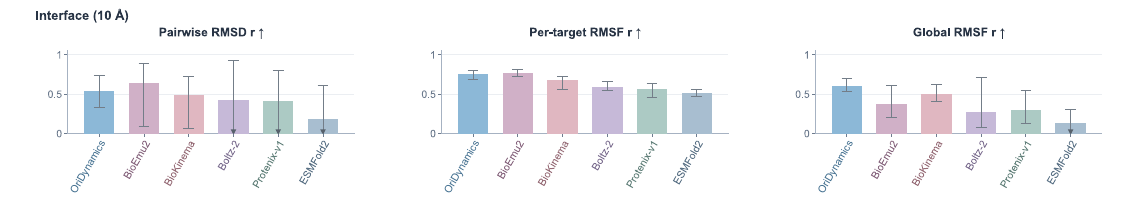}
    \caption[]{\input{results/captions/dynappi_interface_supp}}
    \label{fig:dynappi_interface_correlations}
\end{figure}

%% file: results/captions/dynappi_interface_supp.tex
\textbf{Protein\textendash{}protein interface flexibility on DynaRepo-PPI.} The panels show Pairwise RMSD $r$, Per-target RMSF $r$ and Global RMSF $r$ within fixed 10~\AA{} interfaces. RMSF correlations use heavy atoms. Error bars are 95\% target-bootstrap intervals; downward triangles mark intervals extending below zero.

%% file: application/figure_functional_states_supp.tex
\begin{figure}[!tp]
  \centering
  \begin{minipage}{\linewidth}
  \centering
  \includegraphics[width=\linewidth,keepaspectratio]{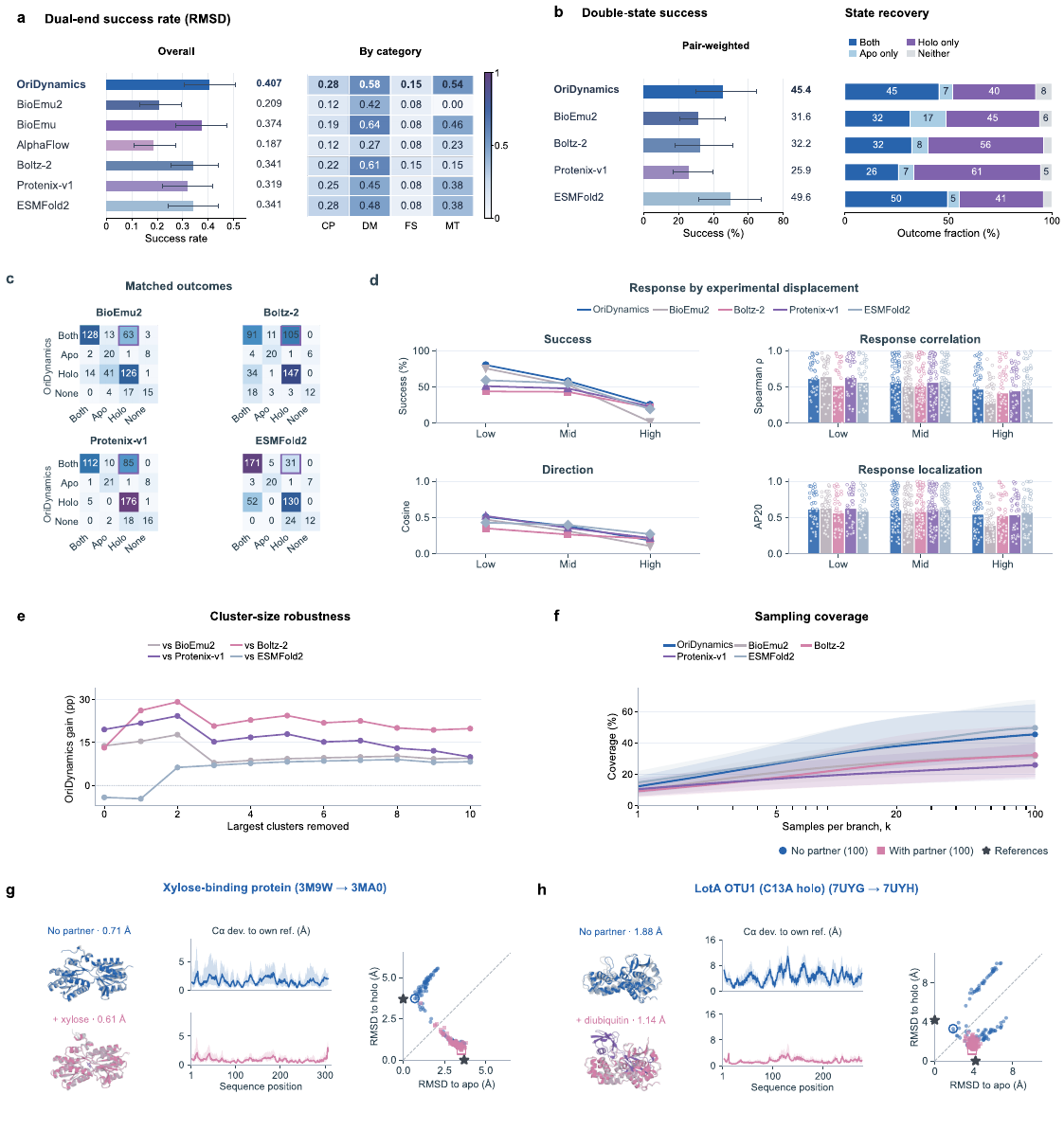}\par
  \caption[]{\input{application/captions/functional_states_supp.tex}}\label{fig:app_functional_states_supp}
  \end{minipage}
  \end{figure}

%% file: application/captions/functional_states_supp.tex
\textbf{Additional analyses of alternative-state and partner-dependent sampling.} \textbf{a,} ConforMix dual-end recovery under the RMSD criterion. \textbf{b,} Pair-weighted ProMiSE double-state success and state-recovery percentages. \textbf{c,} Matched outcome counts against four baselines. \textbf{d,} Cluster-weighted success and response metrics across experimental displacement groups. Response correlations are shown for $\rho\geq0$; bar heights are means over all clusters, including negative correlations. \textbf{e,f,} Sensitivity to removing large clusters and to sampling budget. Error bars and shaded bands in a, b and f show 95\% bootstrap intervals. \textbf{g,h,} Xylose-binding protein and LotA OTU1 (C13A holo reference); blue/pink denote predictions without/with partners. Stars mark references and outlined symbols mark minimum-RMSD representatives. Profiles show median C$\alpha$ deviations to the corresponding reference and 10th\textendash{}90th percentiles. Recovery percentages in stacked bars are rounded to integers.

%% file: application/figure_fm_rm_landscapes_supp.tex
\begin{figure}[H]
\centering
\begin{minipage}{\linewidth}
\centering
\includegraphics[width=\linewidth,keepaspectratio]{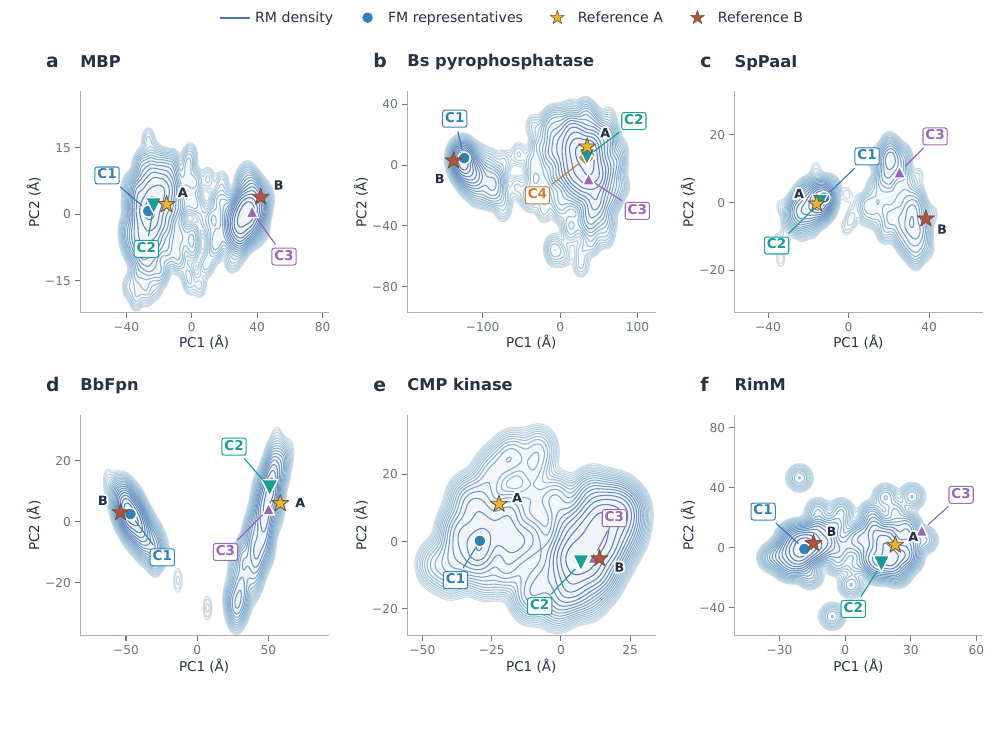}\par
\caption[]{\input{application/captions/fm_rm_landscapes_supp}}\label{fig:app_fm_rm_landscapes}
\end{minipage}
\end{figure}

%% file: application/captions/fm_rm_landscapes_supp.tex
\textbf{FM representatives in refined conformational landscapes.}
\textbf{a\textendash{}f,} Cases follow the order of Fig.~\ref{fig:app_functional_states}c--h.
Blue contours show RM density in an RM-fitted C$\alpha$ PCA space; darker blue denotes higher sampled density.
Colored markers (C1--C4) denote FM cluster representatives; stars mark experimental references A and B.
CMP kinase and RimM pool ensembles generated without and with their respective partners.